\documentclass[longauth]{aa} 

\usepackage{natbib}
\bibpunct{(}{)}{;}{a}{}{,} 
\usepackage{mathtools}
\usepackage{comment}

\usepackage{graphicx}
\usepackage{subfigure}
\usepackage{txfonts}
\usepackage{xcolor}
\usepackage{soul}
\usepackage[colorlinks=True,allcolors=blue]{hyperref}
\newcommand*\ambra{\textcolor{magenta}}

\begin{document}

   \title{The ALPINE-ALMA [CII] Survey: Unveiling the baryon evolution in the ISM of $z\sim5$ star-forming galaxies}

   \subtitle{}

   \author{P. Sawant\thanks{E-mail: prasad.sawant@ncbj.gov.pl}\inst{1}
          \and
          A. Nanni\inst{1,2}
          \and
          M. Romano\inst{3,1,4}
          \and
          D. Donevski\inst{1,5,6}
          \and
          G. Bruzual\inst{7}
          \and
          N. Ysard\inst{8,9}
          \and
          B. C. Lemaux\inst{10,11}
          \and
          H. Inami\inst{12}
          \and
          F. Calura\inst{13}
          \and
          F. Pozzi\inst{14,13}
          \and
          K. Małek\inst{1}
          \and
          Junais\inst{15, 16}
          \and
          M. Boquien\inst{17}
          \and  
          A. L. Faisst\inst{18}
          \and
          M. Hamed\inst{1}
          \and
          M. Ginolfi\inst{19,20}
          \and
          G. Zamorani\inst{13}
          \and
          G. Lorenzon\inst{1}
          \and
          J. Molina\inst{21}
          \and
          S. Bardelli\inst{13}
          \and
          E. Ibar\inst{21}
          \and
          D. Vergani\inst{13}
          \and
          C. Di Cesare\inst{22}
          \and
          M. Béthermin\inst{23,24}
         \and
          D. Burgarella\inst{24}
          \and
           P. Cassata\inst{25,4}
           \and
           M. Dessauges-Zavadsky\inst{26}
           \and
           E. D’Onghia\inst{27}
           \and
           Y. Dubois\inst{28}
           \and
           G. E. Magdis\inst{29,30,31}
            \and
           H. Mendez-Hernandez\inst{32,33}
}

   \institute{National Centre for Nuclear Research, ul. Pasteura 7, 02-093 Warsaw, Poland
   \and
   INAF - Osservatorio astronomico d'Abruzzo, Via Maggini SNC, 64100, Teramo, Italy
   \and
    Max-Planck-Institut für Radioastronomie, Auf dem Hügel 69, 53121 Bonn, Germany
    \and
   INAF - Osservatorio Astronomico di Padova, Vicolo dell'Osservatorio 5, I-35122, Padova, Italy
   \and
   SISSA, Via Bonomea 265, 34136 Trieste, Italy
   \and
   IFPU – Institute for fundamental physics of the Universe, Via Beirut 2, 34014 Trieste, Italy
   \and
   Instituto de Radioastronomía y Astrofísica, UNAM, Campus Morelia, CP 58089 Morelia, Mexico
   \and
   Institut de Recherche en Astrophysique et Planétologie, Université Toulouse III - Paul Sabatier, CNRS, CNES, 9 Av. du colonel Roche, 31028 Toulouse, France
   \and
   Institut d’Astrophysique Spatiale, Université Paris-Saclay, CNRS, 91405 Orsay, France
   \and
   Gemini Observatory, NSF's NOIRLab, 670 N. A'ohoku Place, Hilo, Hawai'i, 96720, USA
   \and
   Department of Physics and Astronomy, University of California, Davis, One Shields Ave., Davis, CA 95616, USA
   \and
   Hiroshima Astrophysical Science Center, Hiroshima University, 1-3-1 Kagamiyama, Higashi-Hiroshima, Hiroshima 739-8526, Japan
   \and
   INAF – Osservatorio di Astrofisica e Scienza dello Spazio di Bologna, Via Gobetti 93/3, 40129 Bologna, Italy
   \and
   Dipartimento di Fisica e Astronomia, Università di Bologna, Via Gobetti 93/2, 40129 Bologna, Italy
   \and
   Instituto de Astrof\'{i}sica de Canarias, V\'{i}a L\'{a}ctea S/N, E-38205 La Laguna, Spain
   \and
   Departamento de Astrof\'{i}sica, Universidad de La Laguna, E-38206 La Laguna, Spain
   \and
   Université Côte d'Azur, Observatoire de la Côte d'Azur, CNRS, Laboratoire Lagrange, F-06000 Nice, France
   \and
   IPAC, California Institute of Technology, 1200 East California Boulevard, Pasadena, CA 91125, USA
   \and
   Dipartimento di Fisica e Astronomia, Università di Firenze, Via G. Sansone 1, 50019, Sesto Fiorentino (Firenze), Italy
   \and
   INAF - Osservatorio Astrofisico di Arcetri, Largo E. Fermi 5, I-50125, Firenze, Italy
   \and
   Instituto de Física y Astronomía, Universidad de Valparaíso, Avda. Gran Bretaña 1111, Valparaíso, Chile
   \and
   Institute of Science and Technology Austria (ISTA), Am Campus 1, 3400 Klosterneuburg, Austria
   \and
   Université de Strasbourg, CNRS, Observatoire Astronomique 392 de Strasbourg, UMR 7550, 67000 Strasbourg, France
   \and
   Aix Marseille Université, CNRS, CNES, LAM, Marseille, France
   \and
   Dipartimento di Fisica e Astronomia, Università di Padova, Vicolo dell’Osservatorio 3, 35122 Padova, Italy
   \and
   Observatoire de Genève, Université de Genève, 51 Ch. des Maillettes, 1290 Versoix, Switzerland
   \and
   University of Wisconsin, 475 N Charter Str., Madison, WI, USA
   \and
   Institut d’Astrophysique de Paris, UMR 7095, CNRS, Sorbonne Université, 98 bis boulevard Arago, 75014 Paris, France
   \and
   Cosmic Dawn Center (DAWN), Jagtvej 128, 2200 Copenhagen N, Denmark
   \and
   DTU-Space, Technical University of Denmark, Elektrovej 327, 2800 Kgs. Lyngby, Denmark
   \and
   Niels Bohr Institute, University of Copenhagen, Jagtvej 128, 2200 Copenhagen N, Denmark
   \and
   Departamento de Astronomía, Universidad de La Serena, La Serena, Chile
   \and
   Instituto de Investigación Multidisciplinar en Ciencia y Tecnología, Universidad de La Serena, La Serena, Chile
   }

   \date{}

\abstract
{Recent observations suggest a significant and rapid build-up of dust in galaxies at high redshift ($z>4$), presenting new challenges to our understanding of galaxy formation in the early Universe. Although our understanding of the phyiscs of dust production and destruction in the galaxies' interstellar medium (ISM) is improving, investigating the baryonic processes in the early universe remains a complex task owing to the inherent degeneracies in cosmological simulations and chemical evolution models.}
{In this work, we characterize the evolution of 98 $z\sim5$ star-forming galaxies observed as part of the ALMA Large Program ALPINE by constraining the physical processes underpinning the gas and dust production, consumption, and destruction in their ISM.}
{We make use of chemical evolution models to simultaneously reproduce the observed dust and gas content of our galaxies, as obtained from spectral energy distribution fitting and ionized carbon measurements, respectively. For each galaxy, we constrain initial gas mass, gas inflows and outflows, and efficiencies of dust growth and destruction. We test these models with both the canonical Chabrier and a top-heavy initial mass function (IMF), the latter allowing for rapid dust production on shorter timescales.}
{We successfully reproduce the gas and dust content in most of the older galaxies ($\gtrsim600$~Myr) regardless of the assumed IMF, predicting dust production primarily through Type II supernovae and no dust growth in the ISM, as well as moderate inflow of primordial gas. In case of intermediate-age galaxies (300 - 600 Myr), we reproduce the gas and dust content through Type II supernovae and dust growth in ISM, though we observe an over-prediction of dust mass in older galaxies, potentially indicating an unaccounted dust destruction mechanism and/or an overestimation of the observed dust masses. The number of young galaxies ($\lesssim$ 300 Myr) reproduced, increases for models assuming top-heavy IMF but with maximal prescriptions of dust production. Galactic outflows are required (up to mass-loading factor of 2) to reproduce the observed gas and dust mass, and to recover the decreasing trend of gas and dust over stellar mass with age. Assuming the Chabrier IMF, models are able to reproduce $\sim$ 65\% of the total sample while, with top-heavy IMF, the fraction increases up to $\sim$ 93\%, alleviating the tension between the observations and the models. Observations from the James Webb Space Telescope will allow us to remove degeneracies in the diverse intrinsic properties of these galaxies (e.g., star-formation histories and metallicity), thereby refining our models.}
{}

   
\maketitle
%
\section{Introduction}
\par The baryon cycle comprises various physical processes that influence the evolution of galaxies throughout cosmic time. The gas within a galaxy's interstellar medium (ISM) can cool, leading to star formation. As stars evolve, they enrich the ISM of galaxies with heavy elements expelled through stellar winds and supernovae (SN) explosions, altering its initial composition. Dust particles form in SN remnants and are primarily composed of silicates and carbonaceous grains. Galactic outflows, dust destruction by SN shock waves, and dust growth processes all impact the ISM's dust and gas content in differing magnitudes, as well as the galaxies' surroundings (e.g., \citealt{Asano13, Christensen18, Casey18, Fujimoto20, Graziani20, Nanni20, Donevski20, Vijayan22, Romano24}). Understanding the baryon cycle is thus crucial for deciphering the evolution of gas, dust and metal content in galaxies (for a review see e.g., \citealt{Tumlinson17,Tacconi20}). 

\par Dust, although constituting only 1\% of the total interstellar matter in the Universe (e.g., \citealt{Tumlinson17,Sarangi18}), plays a pivotal role in shaping the physical and chemical processes that govern galaxy evolution. Notably, it modifies the spectral energy distribution (SED) of galaxies by attenuating the ultraviolet (UV) and optical light from young stars, re-emitting it at longer wavelengths (e.g., \citealt{Calzetti00, Salim20}). This highlights the critical importance of dust in understanding galactic evolution throughout cosmic time.
 
The study of dusty galaxies and, in general, dust in the Universe has witnessed significant progress over the past few decades (e.g., \citealt{Casey14, Heinis14, Wang16, Whitaker17, Zavala18, Williams19, Liu19, Fudamoto20, Gruppioni20, Romano20, Hodge20, Schneider24}). Observational facilities such as \textit{Herschel}, the James Clerk Maxwell Telescope, and the South Pole Telescope have enabled comprehensive studies of dust emission up to $z\sim4$ (e.g, \citealt{Ivison10, Rodighiero11, Combes12,Magnelli13, Lemaux14, Robson14,Bethermin15,Spilker16,Nayyeri17,Donevski20, Kokorev21}). However, the mechanisms responsible for dust formation and growth in galaxies, both locally and at high redshifts, remain a topic of ongoing debate. 

\par The advent of the Atacama Large Millimeter/submillimeter Array (ALMA) has revolutionized our understanding of high-redshift galaxies by enabling systematic detections of cold dust and gas in  $z > 4$ normal\footnote{We refer here to sources lying on the main-sequence of star-forming galaxies, i.e., a tight relation between their star formation rate (SFR) and stellar mass (e.g., \citealt{Speagle14}).} star-forming galaxies. ALMA's increased sensitivity and high spatial resolution have revealed that a significant fraction of star formation activity in the early Universe occurred in heavily dust-enshrouded galaxies \citep{Heinis14, Bouwens16, Whitaker17, Laporte17, Fudamoto20, Gruppioni20, Inami22, Sugahara21, Algera23}. Several studies have focused on the detection of individual high-$z$ targets \citep{Hodge13, Riechers14, Capak15, Watson15, Strandet17}, as well as on the observations of larger areas on the sky \citep{Hezaveh13, Xu14, Dunlop17, Allison19, Jin19, GonzalezLopez20, Pantoni21, Scholtz23}. 

    \par These studies demonstrated that the galaxies which are dusty in nature were observed to dominate the luminous end of the dust luminosity distribution (\citealt{Scoville16, Bethermin17, Donevski18, Dudzeviciute20}). With observations targeted for strong sub-mm sources (\citealt{Wagg12, Carilli13, Riechers13, Riechers14, Fudamoto17, Strandet17, Koprowski20}), there exists a skew towards population studies of starburst systems thus giving rise to the difficulty of identifying a large population of dusty normal star-forming galaxies (DSFGs) at redshifts $> 4$. Consequently, this results in an unconstrained understanding of the evolution of stellar, gas, and dust masses for a statistically significant sample of DSFGs (\citealt{Liu19, Donevski20, Lovell21}). A thorough census of DSFGs is necessary in order to quantify the contribution of these galaxies to the cosmic star formation rate density as well as to understand the early phases of the galaxy formation.

    \par From a theoretical perspective, various attempts have been employed to elucidate the origin and evolution of these galaxies through cosmological simulations (\citealt{Narayanan15, McKinnon17, Dave19, Aoyama19, Hou19, Lovell21, Triani20, Cochrane23, Jones24}) or semi-analytical (\citealt{Lacey16, Popping17, Cousin19, Vijayan19, Lagos19, Pantoni19}) and chemical models (\citealt{Asano13, Calura17, DeLooze20, Nanni20,Pozzi21,Palla24}). These theoretical models concurrently follow the chemical evolution and physical processes in galaxies, which are critical for describing the dust cycle in their ISM. However, despite the diverse approaches and methodologies, these models face challenges in accurately reproducing the baryonic content of high-redshift DSFGs. This suggests the possible onset of more extreme dust production mechanisms, such as larger SN condensation fractions, a significant variation in dust temperature and/or in initial mass functions (IMFs; \citealt{Gall18, Dayal22}).
    
    \par Investigating galactic evolution often involves assuming canonical IMFs (e.g., \citealt{Chabrier03} or \citealt{Kroupa13}) to derive galactic parameters. These IMFs efficiently reproduce galaxies' observables without conflicting with theoretical models in most cases. The emergence of sub-millimeter window on galactic evolution and formation (\citealt{Smail97, Hughes98, Barger98}), revealed that a high amount of star formation occurs in heavily dust-enshrouded galaxies at high redshifts (for review, refer to \citealt{Casey14}). This motivated many studies to investigate the basis of the star formation process, i.e., the IMF.
    
    \par In recent works, a top-heavy IMF (hereafter, THIMF) has been adopted to explain the observational properties of ultra-compact dwarf galaxies (\citealt{Dabringhausen12}), ultra-faint galaxies (\citealt{Geha13, McWilliam13}) and galactic globular clusters (\citealt{Marks12}) in local universe. Also, evidence of THIMF has been reported to account for the low C/O abundance ratio found in local and high-$z$ sub-mm/infrared galaxies (\citealt{Sliwa17, Brown19, Zhang18}). Furthermore, works driven by the James Webb Space Telescope (JWST), have hinted at inconsistencies between observations and models for high-redshift galaxies (\citealt{Boylan23, Labbe23}), suggesting the presence of a THIMF in the high-redshift Universe (\citealt{McKinnon17, Inayoshi22, Sneppen22, Steinhardt22, Bekki23, Steinhardt23, Sun24}). A THIMF, favoring the formation of massive stars, can alleviate the tension between models and the observations as a larger number of SN can lead to rapid chemical enrichment of the ISM corresponding to a rapid dust mass build-up \citep{Palla20}. At the same time, the increased metal availability in the ISM further favors dust growth, necessary for replicating the dust content in high-redshift DSFGs \citep{Algera24}.

    \par Of particular relevance for understanding the dust and gas production and interplay in these galaxies is the ALMA Large Program to INvestigate [CII] at Early times (ALPINE; \citealt{LeFevre20}). This targeted survey observed the singly ionized carbon line (hereafter, [CII]) at 158~$\mu$m and the surrounding dust-continuum emission in a statistically-significant sample of 118 galaxies located at $z\sim 4.4 - 5.9$, when the Universe was 0.9 - 1.5 billion years old. This represents the so-called \textit{early growth phase}, a transition phase between primordial galactic formation \textbf{($z > 6$)} and the onset of the peak of cosmic star formation rate density \textbf{($z \sim 2-3$)}, when galaxies reached their chemical maturity. The ALPINE project has conducted a panchromatic characterization of a hundred of $z\sim5$ star-forming galaxies, providing fundamental information about their morphological and kinematic status (e.g., \citealt{LeFevre20, Jones21, Romano21}), their gas, dust, and metal content (e.g., \citealt{Dessauges20, Gruppioni20, Pozzi21, Vanderhoof22}), or the mechanisms driving their baryon cycle (e.g., \citealt{Fujimoto20,Ginolfi20a,Ginolfi20b}).

    \par In this work, we take advantage of the ALPINE observations to constrain the physical processes governing gas and dust production/consumption in the ISM of post-reionization galaxies. We employ chemical evolution models to reproduce the dust content in these sources reproducing, for the first time, the observed gas content, constraining metallicity, and outflow efficiency in a consistent way, aiming to provide a comprehensive interpretation of their formation and evolution. Furthermore, we test the hypothesis of a non-conventional IMF indicating different channels of dust enrichment in galaxies at high-redshift.

    The paper is structured as follows. In Sect. \ref{sec:data}, we briefly describe the ALPINE sample. The adopted methodology and chemical evolution models utilized to characterize the ISM of high-$z$ galaxies are presented in Sect. \ref{sec:methodology}. We present our results in Sect. \ref{sec:results}, and discuss them in Sect. \ref{sec:discussion}. Finally, in Sect. \ref{sec:conclusions}, we provide a summary of the work and our conclusions.

    \par Throughout this work, we adopt a $\Lambda$CDM cosmology with $H_0 = 70~\text{km s}^{-1}~\text{Mpc}^{-1},~\Omega_m = 0.3~\text{and}~\Omega_{\Lambda} = 0.7$. Furthermore, we use both a \cite{Chabrier03} and a THIMF based on \cite{Larson98}.

\section{Data}\label{sec:data}

The ALPINE sample comprises 118 star-forming galaxies observed in [CII] emission line and far-infrared (FIR) continuum at redshifts $4.4<z<5.9$, excluding the redshift range $4.6<z<5.1$ for which the [CII] line falls within a low-transmission atmospheric window. The targets were originally selected from the Cosmic Evolution Survey (COSMOS; \citealt{Scoville07a,Scoville07b}) and Extended \textit{Chandra} Deep Field South (ECDFS; \citealt{Giavalisco04,Cardamone10}) fields, all possessing accurate spectroscopic redshifts obtained from previous observational campaigns \citep{LeFevre15,Tasca17,Hasinger18}.     

\par These galaxies were selected in the rest-frame UV along the main-sequence of star-forming galaxies at $z\sim5$ (e.g., \citealt{Speagle14}). They exhibit stellar masses within the range $\mathrm{log(M_{*}/M_{\odot}})\sim9-11$, and star-formation rates (SFRs) in the range $\mathrm{log(SFR/M_{\odot}~yr^{-1}})\sim1-3$, as determined through SED fitting \citep{Faisst20} of their multi-wavelength data, encompassing UV to X-ray and radio bands (e.g., \citealt{Hasinger07,Koekemoer07,McCracken12,Guo13,Smolcic17}). 
\begin{table*}[h]
\begin{center}
\setlength{\tabcolsep}{12pt} 
\caption{Parameters for the SED-fitting of the galaxies adopted from \cite{Burgarella22}. The parameters for dust emission are derived from stacked IR template.}
\label{tab:SED_fit_parameter}
\begin{tabular}{lcc}
\hline
\textbf{Parameters}                         & \textbf{Symbol} & \textbf{Range}                             \\
\hline
Mass fraction of PAH               & $q_{PAH}$ & 0.47                              \\
Minimum radiation field            & $U_{min}$ & 17.0                              \\
Power law slope d$U$/d$M_{\mathrm{Dust}}$ $\sim$ $U^{\alpha}$ & $\alpha$ & 2.4                               \\
Dust fraction heated by starlight              & $\gamma$  & 0.54                              \\
\hline
e-folding timescale [Myr]                & $\tau_{main}$ & 500                               \\
Age of main stellar population [Myr]    &   $Age_{main}$ & 101 log values in {[}2 - 1200{]}  \\
Burst                              & $f_{burst}$ & No burst                          \\
Metallicity of the single stellar populations             &   Z     & 0.004                             \\
Ionization parameter               &   $ logU$ & -2.5, -2.0, -1.5                  \\
Color excess                       & $ E\_BV $ & 101 log values in {[}0.01, 1.0{]} \\
Power law slope                    &  $\delta$      & 0.0                               \\
\hline
IMF slope                                &   $\xi$     &  \citealt{Chabrier03} , TH (1.35, 1.5, 1.8)\\
\hline
\end{tabular}
\end{center}
\end{table*}

\par The ALPINE targets were observed for $\sim70$h during Cycles 5 and 6 in ALMA Band 7 (275-373 GHz). Data reduction and calibration were performed using the standard Common Astronomy Software Applications (CASA; \citealt{McMullin07}) pipeline. Each data cube was continuum-subtracted to produce line-only cubes with channel width of $\sim25$~km/s and beam size $\sim1''$ (with a pixel scale of $\sim0.15''$; \citealt{Bethermin20}). A line search algorithm was applied to each continuum-subtracted cube resulting in 75 [CII] detections ($S/N > 3.5$) out of 118 ALPINE targets (including 23 sources detected in continuum). We refer to \cite{LeFevre20}, \cite{Bethermin20}, and \cite{Faisst20} for a detailed description of the ALPINE survey, observation and data processing, and the ancillary data, respectively. All high-level data products are publicly available to the community through ALMA archive and the collaboration website\footnote{A2C2S: \url{https://data.lam.fr/a2c2s/home}}.
\par Following \cite{Burgarella22}, we selected only the ALPINE galaxies with more than 5 data points in the UV-optical coverage, and with $\mathrm{S/N>2.5}$ in each individual rest-frame UV-optical-NIR band. This selection is crucial for retrieving robust constraints on the physical parameters of our galaxies (see Sect. \ref{subsec:SED_fitting}), and to allow for a better description of the gas and dust cycle in their ISM. Therefore, our final sample consists of 98 sources, out of which 68 are detected in [CII]. Further, 21 galaxies are also detected in the dust continuum (of which 19 sources are detected with both the [CII] and the dust continuum). In the following, we will treat both detections and non-detections (either in [CII] and continuum) in the same way, assuming they are drawn from the same galaxy population (see Appendix~\ref{app:upper_limits}).

\section{Methodology}\label{sec:methodology}

    We aim to estimate physical parameters (i.e., stellar mass and SFR) derived from SEDs (Sect.~\ref{subsec:SED_fitting}), dust mass (Sect.~\ref{sec:dust_masses}), and gas mass from [CII] measurements (Sect.~\ref{subsec:gas_mass}), for the sources introduced in Sect.~\ref{sec:data}. In Sect.~\ref{subsec:methodology_dust}, we describe a comprehensive model of the baryon cycle employed to i) constrain parameters dictating gas content of the galaxies (initial gas mass, rate of outflow and inflow), and ii) reproduce observed dust content assuming different prescriptions of dust production and destruction presented in Sect.~\ref{sec:results}.
    
    \subsection{SED fitting and dust emission models}\label{subsec:SED_fitting}
    \par In this work, we maintain consistency between the parameters employed in SED fitting and the chemical evolution model (described in Sect.~\ref{subsec:methodology_dust}). Additionally, we have incorporated a THIMF in the SED-fitting code to test the variability of the IMF in these galaxies, in contrast to previous studies. This approach ensures the alignment of the results from our SED fitting with the chemical models consistently, providing insights into various baryonic components of galaxies across multiple wavelength bands.

    We use the Code Investigating GALaxy Emission (\textsc{cigale}; \citealt{Burgarella05,Noll09,Boquien19}) to model the SEDs of our galaxies. \textsc{cigale} is a modeling and fitting tool that operates on the principle of energy balance between the energy absorbed in the rest-frame UV-NIR part of the total SED of galaxy, and its rest-frame infrared emission, employing Bayesian methods to estimate the physical parameters of galaxies.
    \par Here, we provide a careful treatment of galaxies' observed properties and estimate physical quantities that will be compared with predictions from our chemical models.
    \subsection*{Dust emission}
    \par To model the dust emission in our galaxies, we utilize the composite infrared (IR) template constructed by \cite{Burgarella22} using a sample of 27 ALMA-detected galaxies at $4<z<6$ (including 20 ALPINE galaxies detected in dust continuum from our sample). This template accurately reproduces the characteristics of FIR dust emission in high-$z$ galaxies facilitating robust upper limits on their sub-mm flux densities in the case of non-detections (see Appendix~\ref{app:stacked_template}). In particular, we adopt IR template based on the dust emission models by \cite{Draine14}. As the ALPINE galaxies lack data coverage in the rest-frame near-IR and mid-IR regime, the mass fraction ($q_{PAH}$) of polycyclic aromatic hydrocarbon (PAH) grains is not constrained. Its value is fixed to a minimum equal to 0.47 \citep{Burgarella22} which is consistent with the value derived from relation between $q_{PAH}$ and metallicity for normal DSFGs. Other parameters, such as the minimum value of the radiation field ($U_{min}$), the power-law slope of the distribution of its intensity per dust mass ($\alpha$, with d$U$/d$M_\mathrm{Dust}\propto U^\alpha$), and the fraction of dust heated by starlight with $U>U_{min}$ ($\gamma$) in the molecular clouds, were varied employing the full range of available values by \cite{Burgarella22}.
    \subsection*{Star formation history (SFH) and dust attenuation}
    \par \cite{Burgarella22} investigated different SFHs (constant, delayed with fixed $\tau_{main} = 500$~Myr, and delayed with several $\tau_{main}$, where $\tau_{main}$ is the e-folding time of the main stellar population of the galaxy), finding all of them in agreement with the data. Anyway, a delayed SFH with $\tau_{main} = 500$~Myr and without burst was slightly favored over other SFHs by statistical tests, and hence we adopt the same SFH.
    \par We adopt a modified version of \cite{Calzetti00} law with a varied power-law slope ($\delta$) to account for dust attenuation in ALPINE galaxies. \cite{Boquien22} quantified the effect of different dust attenuation curves for the ALPINE sources, finding that the impact of the choice of the attenuation curves on the estimated physical parameters is limited when SED modeling can be used, with no clear impact on the SFR and only a small systematic effect (limited to $\sim0.3~$dex) for the stellar mass.

   \subsection*{Initial Mass Function}
    Here, we test the variability of the assumed IMF by employing: a canonical Chabrier IMF (\citealt{Chabrier03}) and a THIMF. We implement the THIMF (representing a larger number of short-lived massive stars and a more rapid stellar evolution) in \textsc{cigale} in analogy to \cite{Larson98}. We choose the IMF described by a power-law with a slope $\xi$ such that:
    \begin{equation}
        IMF (m) \propto m^{-\xi} 
    \end{equation} 
with $\xi=1.35, 1.5$, and 1.8. Values of $\xi$ are chosen to cover a range of slopes suggested by observational and theoretical studies \citep{Cappellari12, Martin15, Nanni20, Wang24}.
Both in case of the Chabrier IMF and THIMF we used the following normalization:
\begin{equation}\label{eq:imf_norm}
    \int^{M_{\rm U}}_{\rm M_{\rm L}} m~IMF(m)~dm = 1~M_\odot,
\end{equation}
with $\rm M_{\rm L}=0.1$~M$_\odot$ and $\rm M_{\rm U}=100$~M$_\odot$ as in \citet{Bruzual03}.

    We report all the adopted range of parameters used for the SED fitting in Table \ref{tab:SED_fit_parameter}.

\begin{table*}
\begin{center}
\caption{List of parameters adopted in the simulations of metal and dust evolution described in Sect.~\ref{subsec:methodology_dust}. First tests are run in order to select the reference parameters adopted to run systematic calculations. The stellar mass produced by the end of the simulation is always normalized to 1~M$_\odot$. Chemical species are abbreviated as Olivine: ol, Pyroxene: py, and Carbon: car. $f_{\rm cond}$ is the condensation fraction.} 
\label{Table:models}
\setlength{\tabcolsep}{12pt}
\begin{tabular}{l c c}
\hline

\textbf{Theoretical metal yields} &  & \\
\textbf{Stellar Source} & \textbf{Data set and Denomination}  & \textbf{Mass range in M$_\odot$}\\

\hline
Type II SN &  \citet{LC18}, \citet{Prantzos18} - LC18 &  [13, 120] \\ 

AGB & \citet{Cristallo15} - C15 &  [1, 7] \\

Pop III stars & \citet{Heger10} & [10, 100]  \\

Type Ia SN & \citet{Iwamoto99} &  - \\
\hline
\textbf{Systematic Calculations} & &\\
\hline
 $\tau$ [Myr] & 500 & \\
 IMF  & \citet{Chabrier03} & \\

 IMF (Top-Heavy) & $\propto$ M$^{-\xi}$, $\xi= 1.8$ & \\

 M$_{\rm *, fin}$ [M$_\odot$] & 1 & \\
 M$_{\rm Gas, ini}$ [M$_\odot$] & (2 - 6) $\times M_{\rm star, fin}$ & \\
 $\eta_{out}$ & 0 - 3 & \\
 $\eta_{in}$ & 0 - 10& \\
 M$_{\rm swept}$  [M$_\odot$] & 1535 $n_{SN}^{^{-0.202}}[(Z/Z_{\odot}) + 0.039]^{-0.289}$  & \\
 $\epsilon_{SN}$ & 0.1, 0.5, 1.0 & \\

 SN condensation fraction  & $f_{\rm key,  py}=0.05$, $f_{\rm key,  ol}=0$, $f_{\rm key, car}=0.05$ ($f_{\rm cond} = 5 \%$)& \\
 & $f_{\rm key,  py}=0.10$, $f_{\rm key, ol}=0$, $f_{\rm key, car}=0.10$ ($f_{\rm cond} = 10 \%$) & \\
 & $f_{\rm key,  py}=0.25$, $f_{\rm key,  ol}=0$, $f_{\rm key,  car}=0.25$ ($f_{\rm cond} = 25 \%$) & \\
  & $f_{\rm key,  py}=0.5$, $f_{\rm key,  ol}=0$, $f_{\rm key,  car}=0.5$ ($f_{\rm cond} = 50 \%$) & \\
 & $f_{\rm key,  py}=0.75$, $f_{\rm key,  ol}=0$, $f_{\rm key,  car}=0.5$ ($f_{\rm cond} = 75\%$) & \\
 & $f_{\rm key,  py}=1$, $f_{\rm key,  ol}=0$, $f_{\rm key,  car}=0.5$ ($f_{\rm cond} = 100 \%$)& \\

 AGB condensation fraction & $f_{\rm py}=0.3$, $f_{\rm ol}=0.3$,  $f_{\rm ir}=0.01$, $f_{\rm car}=0.5$ & \\

Dust growth efficiency in the ISM & 0.0, 0.5, 1 & \\
\hline
\end{tabular}
\end{center}
\end{table*}
  
\subsection{Chemical evolution model}\label{subsec:methodology_dust}
The evolution of baryons in the ISM of galaxies, i.e., the gas, metals, and dust, are followed as explained in \citet{Nanni20}. 
\subsection*{Gas evolution}
In short, the evolution of the total gas and of the various gaseous species $i$\footnote{Here, for calculating carbon and silicate dust evolution, we consider Carbon (C), Magnesium (Mg), Silicon Monoxide (SiO) and Water (H$_{2}$O).} are derived by integrating the following differential equations with priors on the e-folding timescale $\tau_{main}$ (assumed same as SFH employed in SED fitting, see Sect.~\ref{subsec:SED_fitting}) and the final stellar mass of the each simulation (M$_{\rm star, fin}$) set equal to 1 $M_{\odot}$: 

\begin{equation}\label{Mgas} 
    \frac{dM_{\rm gas}}{dt}=\frac{dM^{\rm SP}_{\rm gas, ej}}{dt} -SFR - \eta_{out} \times SFR+\eta_{in} \times SFR,
\end{equation}

\begin{equation}\label{MZ}
\begin{multlined}
    \frac{dM_{\rm gas, i}}{dt}=\frac{dM^{\rm SP}_{\rm gas, ej, i}}{dt} -SFR \frac{M_{\rm gas,i}}{M_{\rm gas}} - \eta_{out} \times SFR \frac{M_{\rm gas,i}}{M_{\rm gas}}+\\
    \eta_{in} \frac{M_{\rm gas, prim, i}}{M_{\rm gas, prim}} \times SFR,
\end{multlined}
\end{equation}
where, on the right-hand side, the first two terms of each equation represent the gas ejected by the stellar population at each time and the astration due to star formation, respectively. The two aforementioned terms are calculated with the One-zone Model for the Evolution of GAlaxies (\textsc{omega}) code \citep{Cote17}. The third term of each equation represents the amount of gas ejected from the galaxy through outflow per unit time (namely, the outflow rate, $\dot{M}_{\rm{out}}$), where $\eta_{out}$ is the ``mass-loading'' factor which parameterizes the outflow efficiency as $\eta_{out}\equiv\dot{M}_{\rm{out}}/\rm{SFR}$. Here, we assume that the outflow is driven by stellar feedback and that is $0<\eta_{out}<3$. This is based on the results by \cite{Ginolfi20b}, who found $\eta_{out}\sim1$ from the stacked [CII] spectrum of the ALPINE sources, with a factor 3 of uncertainty after correcting for the contribution of multi-phase gas in the outflow\footnote{In \cite{Ginolfi20b}, it is assumed that [CII] is mostly tracing the atomic gas. If also the ionized and molecular gas are contributing at the same level to the outflow, then $\eta_\mathrm{out, TOT}=3\times \eta_\mathrm{out, [CII]}$.}. Similarly to the outflow, the efficiency of the inflow of gas is parameterized through $\eta_{in}$. The inflowing gas $M_{\rm gas, prim}$ and $M_{\rm gas, prim, i}$ are assumed to be of primordial composition.

\subsection*{Dust evolution}
The metal enrichment from stars in Eq.~\ref{MZ} is fed as input for computing the evolution of each dust species $j$ in the ISM (the molecules and atoms in the gas phase from which each dust species is formed are listed in Table~\ref{Table:models}). This includes dust enrichment from stellar sources, dust destruction from SN explosions, dust growth in the ISM, and outflows:
\begin{equation}\label{eq:dust}
\begin{multlined}
    \frac{dM_{\rm dust, j}}{dt}=\frac{dM^{\rm SP}_{\rm dust, ej, j}}{dt} -SFR \frac{M_{\rm dust, j}}{M_{\rm gas}} - \frac{dM^{\rm SN}_{\rm destr, j}}{dt}- \\-\eta_{out} \times SFR \frac{M_{\rm dust, j}}{M_{\rm gas}}+\frac{dM_{\rm growth, j}}{dt}.
\end{multlined}
\end{equation}
The first term on the right-hand side stands for the dust enrichment with the dust yields approximated as: 
\begin{equation}
\frac{dM^{\rm SP}_{\rm dust, ej, j}}{dt}= \frac{f_{\rm key, j}}{n_{\rm key, j}}\frac{dM^{\rm SP}_{\rm gas, ej}}{dt} \frac{m_{\rm dust,j}}{m_{\rm key, j}},
\end{equation}
where $f_{\rm key, j}$ is the fraction of key element\footnote{We define as ``key element'' the least abundant among the elements that form a certain dust species divided by its number of atoms in the compound \citep{Ferrarotti06}.} locked in the dust as provided in Table~\ref{Table:models} (i.e. condensation fraction). $n_{\rm key, j}$ is the number of atoms of the key element in one monomer of dust, $m_{\rm dust,j}$ is the mass of the dust monomer and $m_{\rm key, j}$ is the atomic mass of the key element.
\par We vary the SN (type Ia and II) condensation fraction of the key element from 5\% to 100\%. The range of selected condensation fractions reflect the uncertainties on dust production for these sources \citep{Marassi19}. 
\par The dust condensation for AGB stars (low and intermediate mass stars going through the asymptotic giant branch phase) provided in Table~\ref{Table:models} is selected on the basis of consistent calculations of dust formation in the circumstellar envelopes of these stars e.g. \citep{Ventura12, Nanni13}. According to \citet{Nanni13}, this value can vary between 0.4 and 0.6 during the superwind phase, when most of the dust is condensed.
\begin{figure*}[h]
  \centering
    \includegraphics[width=0.8\textwidth]{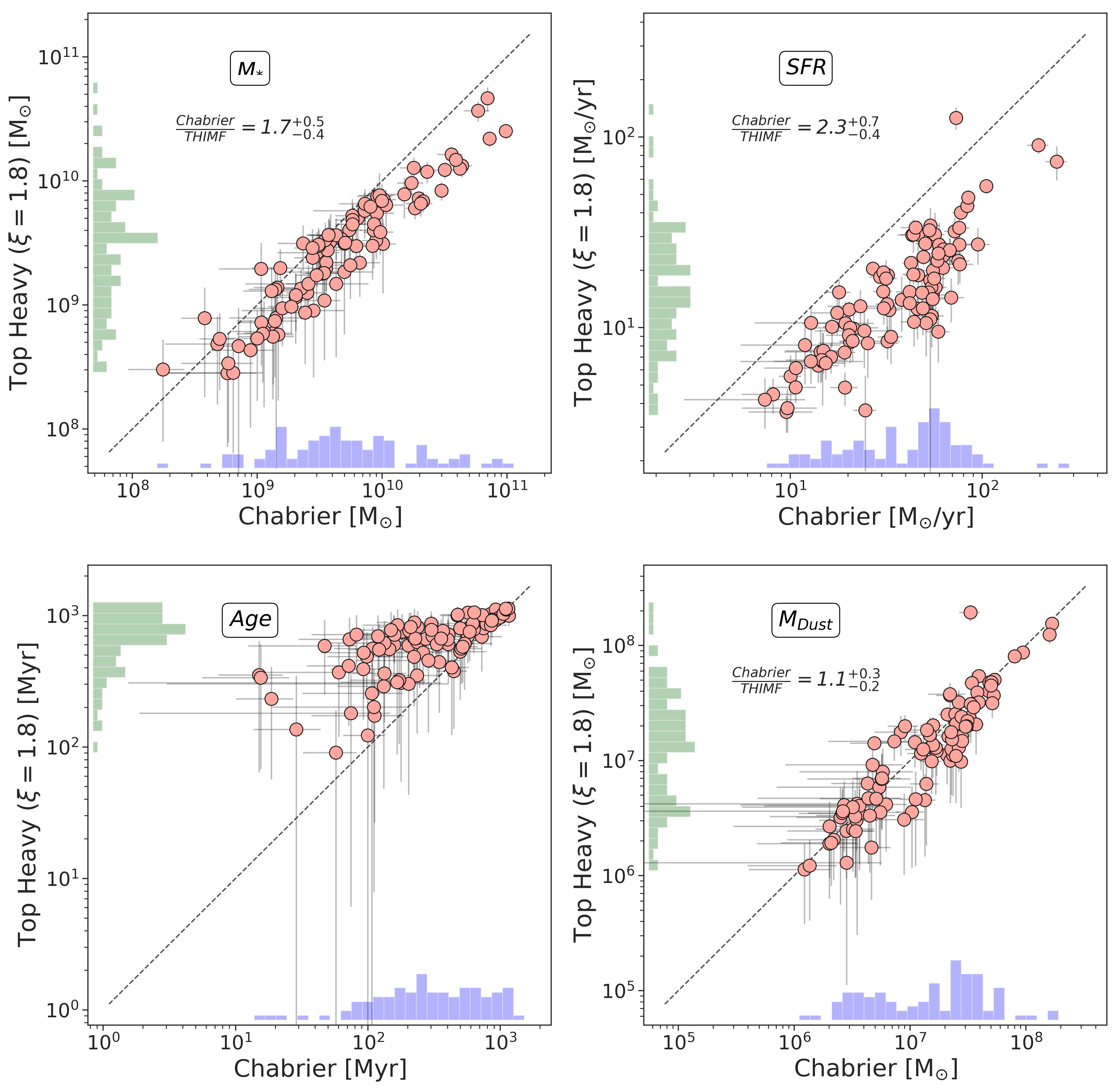}
    \caption{Stellar mass, SFR, age and dust mass (from top-left to bottom-right) plotted for galaxies assuming THIMF against Chabrier IMF. The estimates are derived from SED-fitting code \textsc{cigale} with errors calculated using a bayesian analysis. The histograms on each axis show the distribution of the galaxies. Offsets in $M_{*}$, SFR, and $M_\mathrm{Dust}$ among both IMFs are shown as ratios of the median values (and corresponding uncertainties from the 25th and 75th percentile of the distributions). The black dotted line in each panel represents the 1:1 relation.}\label{fig:sMGas_Age}
    \label{fig:props}
\end{figure*}
The second term of Eq.~\ref{eq:dust} is dust astration due to star formation. The third term represents the destruction of grains operated by SN shocks which is computed through the destruction time-scale, $\tau_{\rm d}$:
\begin{equation}
     \frac{dM^{\rm SN}_{\rm destr, j}}{dt}=\frac{M_{\rm dust, j}}{\tau_{\rm d}},
\end{equation}
with
\begin{equation}
    \tau_{\rm d}=\frac{M_{\rm gas} (t)}{\epsilon_{SN}~R_{\rm SN} (t)~M_{\rm swept}},
\end{equation}
where $M_{\rm gas}(t)$ is the gas mass as a function of time, $M_{\rm swept}$ is the gas mass swept up at each SN event, $\epsilon_{SN} = 0.1-1$ is the destruction efficiency, and $R_{\rm SN}(t)$ is the SN rate, which depends on the SFH and the IMF. We adopt the formalism for $M_{\rm swept}$ from \cite{Asano13} which accounts for the dependence of the swept ISM mass on density and the metallicity of the ISM:

\begin{equation}
    M_{\rm swept} = 1535 ~n_{SN}^{^{-0.202}}[(Z/Z_{\odot}) + 0.039]^{-0.289},
\end{equation}
where $n_{SN}$ = 1.0 $\rm cm^{-3}$ is the gas density around SN Type Ia and II, and $Z_{\odot}$ = 0.0134 \citep{Asplund09}.
We also tested the prescription for M$_{\rm swept}$ from \citet{Priestley22} and adopted in \citet{Calura23}, finding no substantial difference in the results for out best choice of input parameters.
The last term of Eq.~\ref{eq:dust} is dust growth in the ISM. The build-up of dust in the ISM is expressed as:
\begin{equation}\label{eq:dust_growth}
     \frac{dM_{\rm growth, j}}{dt}=4\pi\frac{da_{\rm j}}{dt}a_{\rm j}^2 \rho_{\rm j} n_{\rm s, j},
\end{equation}
where $\rho_{\rm j}$ is the mass density of the dust species $j$, $a_\mathrm{j}$ is the dust size, and $da_{\rm j}/dt$ is the variation of the dust size due to the accretion of atoms and/or molecules on the grain surface and is computed following \citet{Nanni20}. The quantity $n_{\rm s, j}$ is the number of seed nuclei given by the mass of each dust species $j$ divided by the mass of one individual grain \citep{Asano13}. For this calculation we implicitly assume that all the grains in the ISM can potentially act as seed nuclei. We adopt values of 0, 0.5, and 1.0 for the dust growth efficiency, as reported in Table~\ref{Table:models}. The dust growth efficiency is modulated through the term $da_{\rm j}/dt$ which includes the probability for molecules or atoms forming the dust species to stick on the grain surface when they collide with it (sticking coefficient). The sticking coefficient varies from 0 (no molecules or atoms sticking, hence no dust growth) to 1 (all the molecules or atoms are sticking on the grain surface, therefore accretion is fully efficient). We neglect the effect of photo-evaporation of dust grains, which has been shown to be negligible irrespective of the initial gas mass, SFH and IMF \citep{Nanni24}. We selected two cases for the IMF of stars, consistently with the SED fitting, as described in Sect.~\ref{subsec:SED_fitting}.
The values adopted for each parameter of the models are reported in Table \ref{Table:models}.

\subsection*{Dust luminosity}\label{sub:dust_lum}
For each galaxy, assuming the fixed parameters for dust emission model from the SED fitting procedure for consistency (i.e., $U_{\rm min}=17$, $\alpha=2.4$, $\gamma=0.54$, and $q_{PAH}=0.47$) in Table~\ref{tab:SED_fit_parameter}, we compute the dust luminosity as in \citet{Draine07}. We neglected the contribution of PAHs at these longer wavelengths and consider the equilibrium temperature for dust grains for dust species. The emission at a given frequency, $\nu$, for the dust species $j$ (either carbon or silicate) is given by:
\begin{equation}\label{eq: dust_lum}
    L_{\nu, j}= (1-\gamma) L_{\nu, j}(U_{\rm min})+\gamma L_{\nu, j}(U_{\rm min}, U_{\rm max}),
\end{equation}
where $L_{\nu, j}(U_{\rm min})$ is the luminosity of the diffuse dust illuminated by a radiation field scaled by the interstellar radiation field (ISRF) as $U_{\rm min}\times ISRF$ \citep{Mathis77}, $L_{\nu, j}(U_{\rm min}, U_{\rm max})$ is the dust emission in HII regions when illuminated by a variable radiation field of values between $U_{\rm min}$ and $U_{\rm max}=10^7$, and $\gamma$ parameterizes the amount of dust exposed to the radiation field of HII regions.
The diffuse dust emission can be written as:
\begin{equation}
L_{\nu, j}(U_{\rm min})=4\pi M_{{\rm Dust}, j} \kappa_j B(T_{{\rm Dust}, j}(U_{\rm min})),    
\end{equation}
where $M_{{\rm Dust}, j}, \kappa_j$ and $T_{{\rm Dust},j}$ are dust mass, absorption coefficient for each dust species and dust temperature computed for the radiation field $U_{\rm min}\times ISRF$.
The dust emission in HII regions reads:
\begin{equation}\label{eq: dust_lum1}
\begin{multlined}
    L_{\nu, j}(U_{\rm min}, U_{\rm max})= 4\pi M_{{\rm Dust}, j} \kappa_j \frac{\alpha-1}{U^{1-\alpha}_{\rm min}-U^{1-\alpha}_{\rm max}}
    \int_{U_{\rm min}}^{U_{\rm max}}B(T_{{\rm dust}, j}(U))\\ U^{-\alpha} dU.    
\end{multlined}
\end{equation}
The quantity $\kappa_j$ is computed for carbon and silicates starting from the optical absorption coefficient provided by \citet{Draine07} and integrated over the grain size distribution as in \citet{WD01, Draine07}. We finally sum up the contribution of silicates and carbonaceous grains to the dust luminosity:
\begin{equation}
    L_{\nu}=\sum_j L_{\nu, j}.
\end{equation}
Given the total luminosity of a source at a $\nu$ in optically thin regime the total dust mass can be derived as:
\begin{equation}\label{eq:dust_mass}
    M_{\rm Dust}=\frac{L_{\nu}}{4\pi \kappa_{\rm av} L^\prime},
\end{equation}
where
\begin{equation}
\begin{multlined}
    L^\prime= \gamma B(T_{{\rm Dust}, j}(U_{\rm min}))+\\(1-\gamma) \frac{\alpha-1}{U^{1-\alpha}_{\rm min}-U^{1-\alpha}_{\rm max}}
    \int_{U_{\rm min}}^{U_{\rm max}}B(T_{{\rm Dust}, j}(U)) U^{-\alpha} dU,
\end{multlined}
\end{equation}
and
\begin{equation}\label{eq:kappa_av}
    \kappa_{\rm av}=\sum_j \frac{M_{{\rm Dust}, j}\kappa_j}{M_{\rm Dust}}.
\end{equation}

    \subsection{Estimates of the dust masses}\label{sec:dust_masses}

    We estimate the dust mass for each source through the luminosity at 160~$\mu$m derived from the SED fitting by using Eq.~\ref{eq:dust_mass}. From our dust evolutionary models (Sect.~\ref{subsec:methodology_dust}), the predicted dust composition typically consists of $70$\% carbon and $30$\% silicates which we adopt as dust composition in Eq.~\ref{eq:kappa_av}. In contrast, the dust models by \citet{Draine14}, also incorporated in \textsc{cigale} assume a composition of $25$\% carbon and $75$\% silicates for the dust in the diffuse ISM (see Sect. 3.2 in \citealt{Draine14}). Furthermore, we adopt the same optical properties as in \citet{Draine07} which results in a higher dust opacity than that of \cite{Draine14}. Due to this differences in the dust composition and optical properties, our predicted dust masses are $\sim 2$ times lower than those estimated by \textsc{cigale}\ambra{\footnote{We assume a typical density for silicates of 3.1~g~cm$^{-3}$ and of 1.8~g~cm$^{-3}$ for carbonaceous grains.}}. To account for this discrepancy, we divide the dust masses derived from \textsc{cigale} by a factor of 2 for subsequent analysis. We emphasize that the estimates of the dust masses can vary approximately a factor of $\sim$3 with respect to the values found by the SED fitting employing \citet{Draine14} depending on the selected dust absorption coefficient and on the method adopted to perform the SED fitting \citep{Burgarella22}.

\subsection{Estimates of the total gas mass}\label{subsec:gas_mass}
Following \cite{Dessauges20}, we utilize the prescription by \cite{Zanella18} to estimate the molecular gas mass ($\mathrm{M_{Gas}}$) from [CII] luminosity ($L_{\mathrm{[CII]}}$) as
    \begin{equation}\label{eq:mgas}
    \mathrm{log_{10}}(L_{\mathrm{[CII]}}/L_{\odot}) = (-1.28\pm0.21) + (0.98\pm0.02)~\mathrm{log_{10}}(M_{\mathrm{gas}}/M_{\odot}),
    \end{equation}
where $L_{\mathrm{[CII]}}$ is the [CII] luminosity of each galaxy computed by \cite{Bethermin20}. Indeed, \cite{Dessauges20} demonstrated that these measurements can be considered as a reliable proxy for the \textit{total} mass of gas (i.e., including the molecular, atomic, and ionized phase).
 
\subsection{Estimation of optimal model parameters}\label{subsec:best_fit}
We generate a set of evolutionary tracks covering a range of input parameters (initial gas mass in the ISM, condensation fraction of SN, efficiency of galactic inflow and outflow, dust growth efficiency in the ISM; see Table~\ref{Table:models}), in order to derive their most probable values to be used in the description of the baryon cycle in our galaxies. The high degeneracy between SN condensation fraction, outflow efficiency, and dust growth, prevents us from associating them with an optimal value for our models. Therefore, we let these quantities vary in the intervals reported in Table~\ref{Table:models}. For all the other parameters, we determine optimal values for each model that best reproduce our galaxies. We minimize the reduced chi-square for each source ($\chi^2_{\rm gal,m}$) by finding the residual between the "$n^{th}$" predicted galaxy's physical parameter (i.e., $M_\mathrm{gas}$, dust luminosity at 160~$\mu$m, SFR, and age; all of them normalized to the stellar mass) derived from the "$m^{th}$" model, $\widetilde{f}_{n,m}$, and the estimated ones from the \textsc{cigale} SED fitting and observations, $f_{n}$s, along with their uncertainties, $f_{\rm err, n}$:
\begin{equation}\label{chi2}
    \chi^2_{\rm gal, m}= \frac{1}{N}\sum\limits_{\rm n=1}^N \frac{(f_{n}-\widetilde{f}_{n, m})^2}{f_{\rm err, n}^2},
\end{equation}
where $N$ is the total number of physical parameters.

We further derive the mean value $(\overline{f}_{n})$ averaged on all the $M$ models (listed in Table~\ref{Table:model_parameters}) as:
\begin{equation}
    \overline{f}_{n} = \frac{\sum_{\rm m=1}^M \widetilde{f}_{n, m} \times p_{\rm m}}{\sum_{\rm m=1}^M p_{\rm m}},
\end{equation}

where $p_{m}$ is the $\chi^2$ probability of each model "$m$" for the physical parameter "$n$" obtained as (e.g., \citealt{Johnson95}): 
\begin{equation}
    p_{\rm m} =  \frac{1}{2^{N/2} \Gamma(N/2)}~ {\chi^2}_{\rm gal,m}^{(\frac{N}{2}-1)}~ {\rm e^{-\frac{\chi^2_{\rm gal, m}}{2}}}. 
\end{equation} 
We show a schematic representation of the above-described procedure in Appendix~\ref{app: fitting_scheme}.

\begin{figure}[h!]
  \centering
    \includegraphics[width=0.45\textwidth]{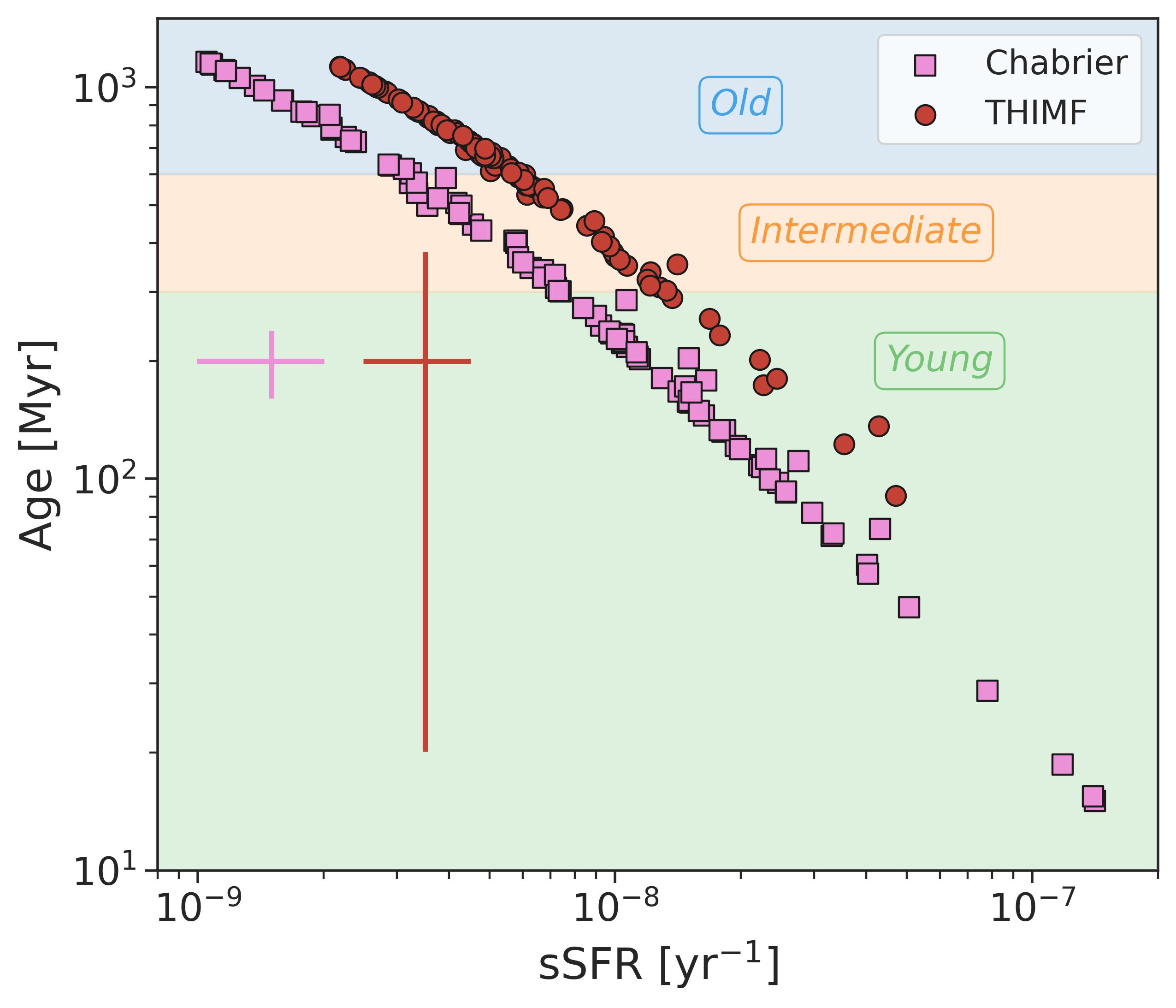}
    \caption{Age of the ALPINE galaxies as a function of sSFR assuming a Chabrier IMF (pink squares) and a THIMF (red circles). Cyan, orange, and green regions represent old ($\gtrsim600$~Myr), intermediate ($300 - 600~$Myr), and young ($\lesssim300$~Myr) population of galaxies. Crossbars indicate the average error on the age and sSFR for the respective sample.}
    \label{fig:sSFR_age}
\end{figure}

\begin{table}[h!]
\centering
\caption{Mean optimal values for the input parameters of our chemical models.}
\label{Table:model_parameters}
\setlength{\tabcolsep}{12pt}

\begin{tabular}[width=0.95\textwidth]{l c}
\hline
\textbf{Model parameter} & $\overline{f}_{n}$ \\
\hline
M$_{\rm Gas, ini}$ [M$_\odot$] (Chabrier) &  3.3 $\times M_{\rm star, fin}$  \\
M$_{\rm Gas, ini}$ [M$_\odot$] (THIMF) & 4.4 $\times M_{\rm star, fin}$  \\
\textbf{$\eta_{in}$} & 0.6 \\
$\epsilon_{SN}$ & 0.1 \\
\hline
\end{tabular}
\label{Table:reactions}
\end{table}

\begin{figure*}[h]
  \centering
    \includegraphics[width=0.9\textwidth]{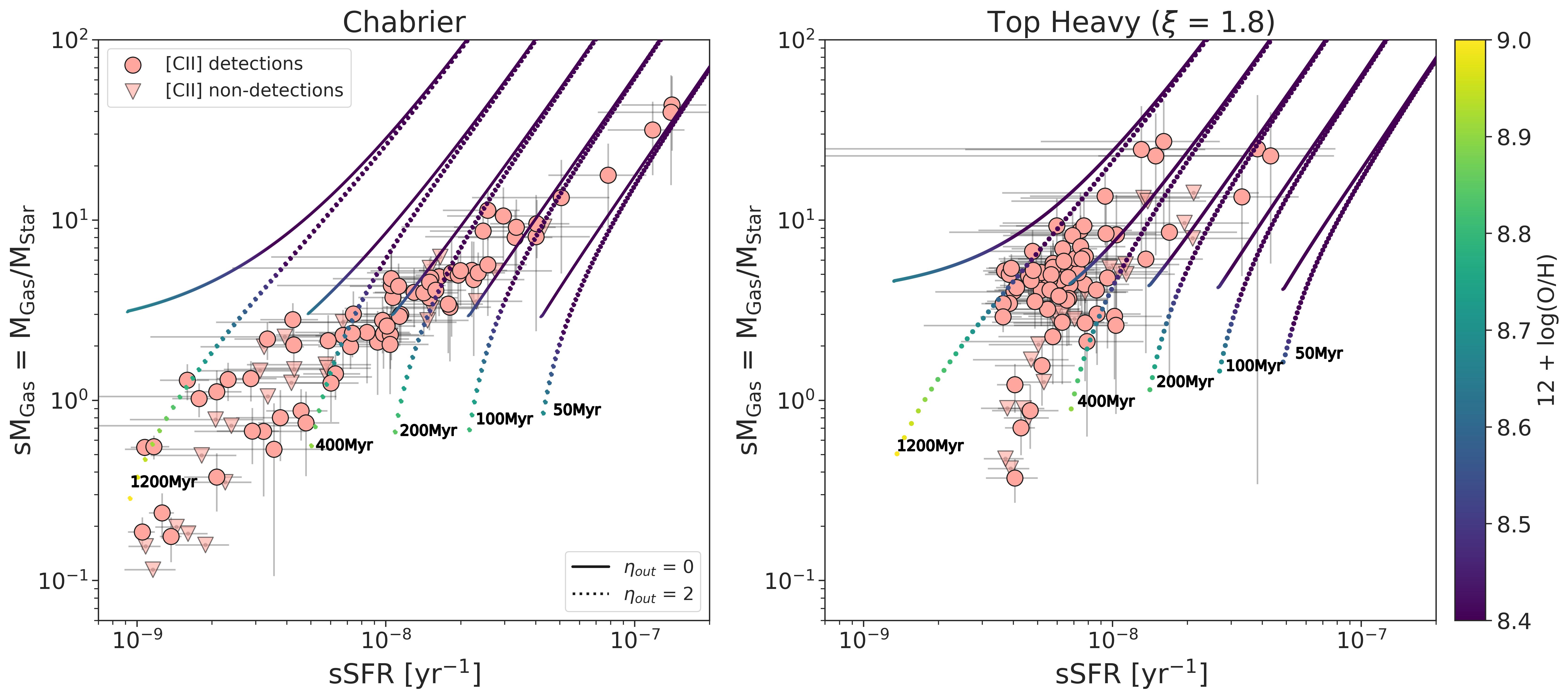}
    \caption{$\mathrm{sM_{Gas}}$ is plotted against sSFR for the ALPINE galaxies assuming a Chabrier (left) and THIMF (right). Circles represent galaxies detected via [CII], while upper limits on gas mass are shown as inverse triangles. Evolutionary models (solid and dotted lines) are plotted for different final ages (increasing from right to left). The figure shows the case with [M$_{\mathrm{Gas, ini}}$, $\eta_{out}$, $\eta_{in}$, $\epsilon_{SN}$] $\sim$ [3.3, 0-2, 0.6, 0.1] and [4.4, 0-2, 0.6, 0.1] for Chabrier and THIMF, respectively. All models are color-coded by the predicted metallicity and illustrate the effect of varying $\eta_{out}$ from 0 (solid curves) to 2 (dotted curves)}.\label{fig:sMGas_Age}
    \label{fig:DFRD_chabrier}
\end{figure*}

\section{Results}\label{sec:results}

 \subsection{SED fitting with different IMFs}\label{subsec:SED_IMF}
    
    \par As described in Sect.~\ref{subsec:SED_fitting}, assuming a delayed SFH with $\tau_{main} =$ 500~Myr and a modified attenuation law based on \cite{Calzetti00}, we obtain estimates for the physical parameters of the galaxies.

    In Fig.~\ref{fig:props}, we compare the distributions of $M_{*}$, SFR, Age, and $M_{\text{Dust}}$ obtained with \textsc{cigale} for both the Chabrier and THIMF. In case of the THIMF, we employ the same SFH and dust emission models as used in the Chabrier IMF to assess the influence of a shallower IMF slope on the derived physical parameters and models. We observed that the physical properties of our galaxies, as derived from SED fitting, were not well constrained by \textsc{cigale} when using THIMF slopes of $\xi = 1.35$ or $\xi = 1.5$, resulting in skewed or overly broad probability distribution functions (PDFs). In contrast, a slope of $\xi$ = 1.8 provided a significantly better fit to the data, as indicated by lower chi-square values and more peaked PDFs. Based on these results, we adopt a fixed THIMF slope of $\xi$ = 1.8 throughout this paper. In the case of Chabrier, our median estimate for stellar mass is $\log_{10}(M_{*}/M_{\odot}) = 9.57_{-0.29}^{+0.38}$, consistent with the estimates from \cite{Faisst20} for the same sample of galaxies. The median value for star-formation rate\footnote{Throughout the paper, we adopt the instantaneous SFR from \textsc{cigale} (see \citealt{Boquien19} for more details), which allows us to make a better comparison with SFR predicted by models at each time-step.} is $\log_{10}(\text{SFR}/M_{\odot}~\text{yr}^{-1}) = 1.63_{-0.33}^{+0.11}$, also consistent with estimates by \cite{Faisst20}. The median dust mass estimate is $\log_{10}(M_{\text{Dust}}/M_{\odot}) = 7.2_{-0.51}^{+0.27}$, consistent with previous studies from \cite{Burgarella22} and \cite{Sommovigo22}, but $\sim 0.7$ dex lower than the estimates from the fiducial model of \cite{Pozzi21}, where lower dust temperatures are assumed ($T_{\text{Dust}} \sim 25$~K)\footnote{\cite{Pozzi21} also assumed higher dust temperature $\sim 35$~K, which reduced their dust estimates by 60\% with respect to their fiducial model, resulting in a better agreement with this work.} as compared to typical values for normal SFGs at $z > 4$ (i.e., $T_{\text{Dust}} \sim 40-60$~K; \citealt{Bakx21,Burgarella22,Sommovigo22}). Furthermore, the median estimate for stellar mass assuming a THIMF is $\log_{10}(M_{*}/M_{\odot}) = 9.47_{-0.39}^{+0.29}$, for star formation rate is $\log_{10}(\text{SFR}/M_{\odot}~\text{yr}^{-1}) = 1.17_{-0.22}^{+0.21}$, and for dust mass is $\log_{10}(M_{\text{Dust}}/M_{\odot}) = 7.12_{-0.49}^{+0.26}$. We note that all our estimates report uncertainties based on the 25th and 75th percentiles of the corresponding distributions.

    \par From Fig.~\ref{fig:props}, we observe that the stellar mass estimates are, on average, a factor $1.7_{-0.4}^{+0.5}$ lower when assuming a THIMF than a Chabrier IMF. Similarly, the THIMF estimates for SFR are lower by a factor of $2.3_{-0.4}^{+0.7}$ than SFR derived assuming Chabrier IMF. This behavior is comparable with previous results from the literature (e.g., \citealt{Wang24}). In case of Chabrier IMF, the predicted ages of the main stellar population of our galaxies lie on the younger end of the distribution (with mean value $\sim$ 350~Myr), whereas for the top-heavy case, the predicted ages lie on the older end (with mean value $\sim$ 650~Myr). In particular, we note that the younger galaxies in our sample tend to deviate more from the 1:1 relation, as further discussed in Sect.~\ref{subsec:dust}. Finally, we do not observe any significant difference in the dust mass between the two assumed IMFs.
    
    \par For a better comparison between observations and predictions, we divide our sample into three arbitrary age bins, i.e., old ($\gtrsim$ 600 Myr), intermediate (300 - 600 Myr) and young galaxies ($\lesssim$ 300 Myr). These are shown in Fig.~\ref{fig:sSFR_age}, together with the relation between age and sSFR in our sources for the Chabrier and THIMF. In both cases, we obtain an almost linear relation between age and sSFR (see also e.g., \citealt{Nanni20,Galliano21,Burgarella22}), strengthening our comparison between evolutionary tracks and observations throughout the paper.
    \par It should be noted that the estimates of the physical parameters are subject to the choice of the SFH (\citealt{Leja19, Lower20, Whitler23}) and the attenuation law (\citealt{Buat19, Hamed23}). For instance, \cite{Topping22} analyzed a sample of 40 UV-bright galaxies at $z\sim7-8$ as part of the Reionization Era Bright Emission Line Survey (REBELS; \citealt{Bouwens22}). They found that the inferred stellar masses and sSFR for main-sequence galaxies in the epoch of reionization can be largely underestimated (up to an order of magnitude) if assuming a constant SFH instead of a non-parametric one. This effect is most prominent in the youngest galaxies with large sSFR and age less than 10~Myr, for which the emission from older stellar populations can be outshined by the presence of more recent bursts.

    \subsection{Constraining the gas content}\label{subsec:specific_gas_mass}
    In this section, we use our models to reproduce the gas content of our galaxies which is dictated by various physical quantities, primarily the initial gas mass ($M_{\mathrm{Gas, ini}}$), the mass-loading factor ($\eta_{out}$) and the inflow parameter ($\eta_{in}$). Observational properties from the literature (i.e., gas mass, metallicity and outflow efficiency) for the ALPINE galaxies assist us to constrain the priors for our models described in Sect. \ref{subsec:methodology_dust}.
    \par We begin by reproducing the observed gas content in the ISM of our galaxies, i.e., 68 ALPINE galaxies detected in [CII] and 30 [CII] non-detections (see Sect. \ref{sec:data}). Specifically, we assign an initial mass of gas to our galaxies that is proportional to their final stellar mass (normalized to 1~M$_\odot$; see Table~\ref{Table:models}). This is a crucial parameter as it drives the evolution of galaxies over cosmic time, and it is deeply linked with their metallicity history as well as with the outflow and inflow of gas. 

    As mentioned in Sect. \ref{subsec:methodology_dust}, we test values of $\eta_{out}$ between 0 and 3, based on observations by \citealt{Ginolfi20b}. Regarding the gas-phase metallicity, given the stellar mass of our $z\sim5$ galaxies, we expect them to be characterized by a sub-solar metallicity (e.g., \citealt{Faisst16}). Similar conclusions were reached by \cite{Vanderhoof22}, who measured metallicity from absorption lines in the rest-frame UV stacked spectrum of 10 ALPINE galaxies. They found an average metallicity of $\mathrm{12 + 1og(O/H)} = 8.4^{+0.3}_{-0.5}$, corresponding to $\sim50\%$ of solar metallicity. We adopt the value $\mathrm{12 + 1og(O/H)} \sim 9.0$ to constrain the terminal metallicity of the models but it should be noted that the observational value represents a small sub-sample of the ALPINE galaxies. We also varied the inflow parameter $\eta_{in}$ in the calculations between 0 and 10.

     \par We derive the best value of initial gas mass, inflow parameter, and SN destruction efficiency from the best-fitting methods described in Sect.~\ref{subsec:best_fit} and reported in Table~\ref{Table:model_parameters}. We then examine the variation of the outflow efficiency ($\eta_{out}$) on the corresponding evolutionary tracks. Fig.~\ref{fig:sMGas_Age} presents the result of this analysis. We present the observed specific mass of gas (i.e., $\mathrm{sM_{Gas}} \equiv \mathrm{M_{Gas}/M_{*}}$) as a function of specific star formation rate (i.e., sSFR $\equiv \mathrm{SFR/M_{*}}$), assuming both a Chabrier (left) and a THIMF (right), and compare these with our best predictions from models. The uncertainties in $\mathrm{sM_{Gas}}$ are calculated by propagating the errors from $\mathrm{M_{Gas}}$ (from Eq. \ref{eq:mgas}) and $M_{*}$ (from the SED fitting). Each curve illustrates the variation of gas-phase metallicity with age, and the effect of different outflow efficiency on the history of $\mathrm{sM_{Gas}}$. We run the models for different final ages spanning the age of stellar population of the galaxies.
\begin{figure*}[h]
  \centering
    \includegraphics[width=0.9\textwidth]{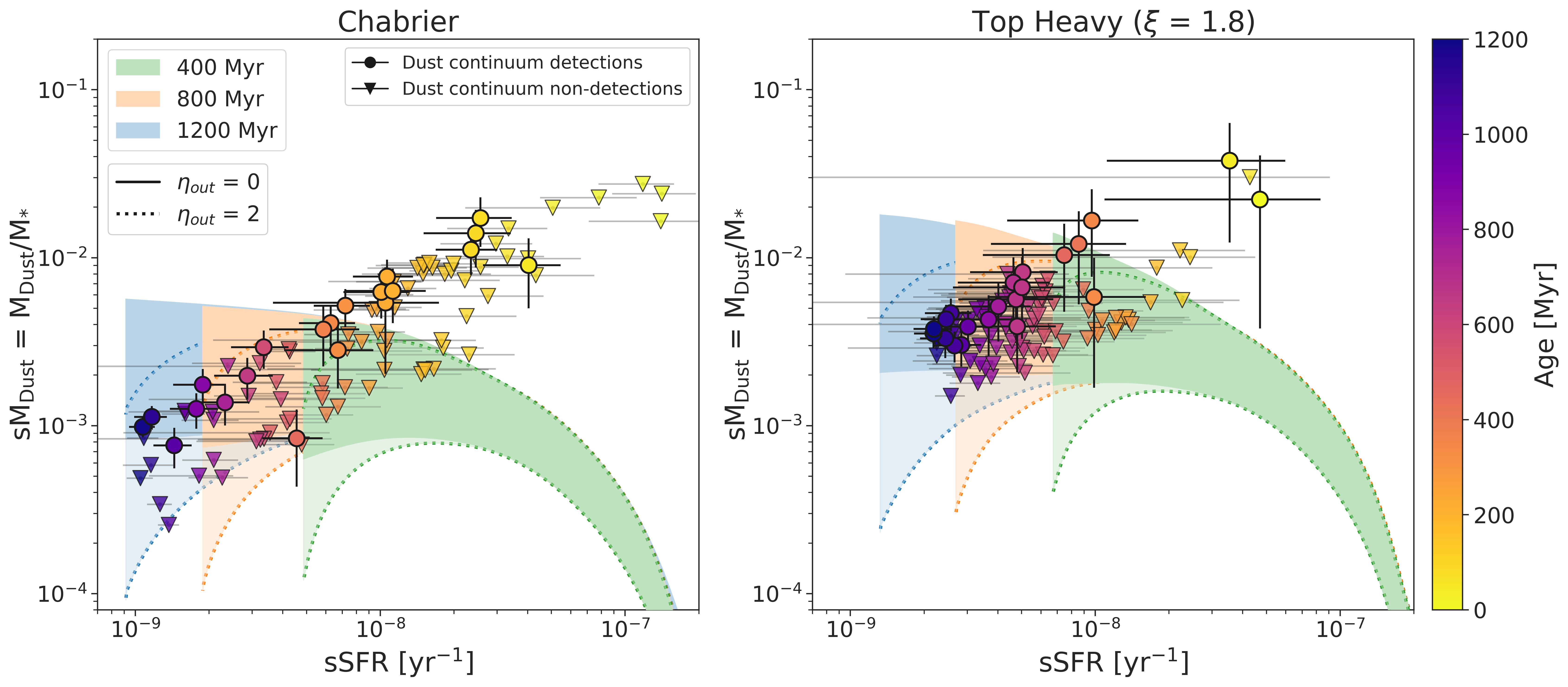}
    \caption{$\mathrm{sM_{Dust}}$ is plotted against $\mathrm{sSFR}$ for the ALPINE galaxies assuming a Chabrier IMF (left panel) and THIMF (right panel). Circles indicate galaxies with detected dust continuum, while inverse triangles represent upper limits on the dust mass. Dust evolution tracks are plotted for the constrained parameters [M$_{\mathrm{Gas, ini}}$, $\eta_{out}$, $\eta_{in}$, $\epsilon_{SN}$] $\sim$ [3.3, 0-2, 0.6, 0.1] and [4.4, 0-2, 0.6, 0.1] for the Chabrier and THIMF models, respectively, and for galaxy ages of 400 Myr, 800 Myr, and 1200 Myr for illustrative purpose. Solid color lines represent models assuming $\eta_{out} = 0$, while dotted lines show the effect of increasing the outflow efficiency to $\eta_{out} = 2$. The models range from 25\% dust condensation without dust growth in the ISM to 100\% condensation with maximum dust growth efficiency. The galaxies are color-coded according to the age of their main stellar population.}
    \label{fig:DFRD_chabrier}
\end{figure*}
    \par Assuming a Chabrier IMF, we reproduce the $\mathrm{sM_{Gas}}$ evolution in all galaxies by adopting the optimal best-fit values $M_{\mathrm{Gas, ini}}=3.3$, $\eta_{in}=0.6$, $\epsilon_{SN}=0.1$ from Table~\ref{Table:model_parameters}, and allowing the mass-loading factor to reach $\eta_{out}=2$. A value of $\eta_{in}=0.6$ is necessary to reproduce the galaxies with lower observed specific gas mass within the adopted constraints of having metallicity $12 + \mathrm{log(O/H)} <9.0$. We anticipate solar to sub-solar metallicities for all of our sources, except for the oldest galaxies (corresponding to the lowest sSFR) for which an almost super-solar metallicity is required (12 + log(O/H) $\sim$ 9.0). From Fig.~\ref{fig:sMGas_Age}, it is clear that star formation alone (solid lines of each set of models) is not enough to deplete the gas content in the ISM of galaxies with age $\gtrsim300~$Myr (i.e., $sSFR\lesssim 10^{-8}~yr^{-1}$) galaxies. Instead, outflows play a significant role, with larger $\eta_{out}$ needed to remove the gas more rapidly than in the $\eta_{out} = 0$ case. In case of Chabrier IMF, we require $\eta_{out}\sim 2$ to successfully reproduce the observed amount of gas in older galaxies, albeit metallicity increases rapidly towards lower sSFR. Assuming a THIMF, we are able to reproduce the $\mathrm{sM_{Gas}}$ evolution in all galaxies but those with lowest values of $\mathrm{sM_{Gas}}$ by adopting $M_{\mathrm{Gas, ini}}=4.4$, $\eta_{in}=0.6$, and letting the outflow efficiency to reach $\eta_{out}=2$. 
    \par Although a larger mass-loading factor could be adopted in this case (i.e., $\eta_{out}>2$, especially for the older sources), it would quickly exceed the observed metallicity constraints \citep{Vanderhoof22}, as well as the observed upper limits on the outflow efficiency \citep{Ginolfi20b}. Hence, we limit our models to $\eta_{out} = 2$.

    \subsection{Reproducing the specific dust mass}\label{subsec:dust}
    The specific dust mass ($\mathrm{sM_{Dust}= M_{Dust}/M_{*}}$), as identified by e.g., \cite{RemyRuyer15,Calura17,Nanni20}, serves as a valuable metric for quantifying the dust content in galaxies as it traces the rate of dust production and/or destruction in their ISM, and can be used to disentangle the main-sequence and starburst populations \citep{Donevski20,Kokorev21}. The evolution of $\mathrm{sM_{Dust}}$ with the sSFR (or age) is also known as dust formation rate diagram (DFRD; e.g., \citealt{Burgarella22}). Analyzing the DFRD provides insights into various phases of the dust cycle, including formation, destruction and transport \citep{daCunha15,  Michalowski19, Nanni20, Whitaker21, Burgarella22, Shivaei22, Donevski23, Lorenzon24}.

    \par To model dust evolution, we based our calculations on the models described in Sect.~\ref{subsec:methodology_dust} and we used as input parameters $M_{\rm Gas, ini}$ and $\eta_{in}$, as derived in Sect.~\ref{subsec:best_fit}. We determine the dust mass from the best-fitted SED using \textsc{cigale}, as described in Sect.~\ref{subsec:SED_fitting} and ~\ref{sec:dust_masses}. The SFR and stellar mass are also computed from the SED fitting following the same procedure as in \cite{Burgarella22}. To model the $\mathrm{M_{Dust}}$, we adopt the state of the art theoretical metal yields mentioned in Table~\ref{Table:models}. The selected yields for Type II SN are from \citet{LC18} and have been shown to be among those providing the highest amount of dust at the beginning of the baryon cycle \citep{Nanni20}. The final metal yields adopted in the models are weighted for the star rotational velocities as in \citet{Prantzos18}. 

    \par In Fig.~\ref{fig:DFRD_chabrier}, we plot the DFRD for the ALPINE galaxies assuming Chabrier IMF (left) and THIMF (right). With variations in stellar mass and SFR as described in Sect.~\ref{subsec:SED_IMF}, compared to the case of Chabrier IMF (mean value for $\mathrm{sSFR_{Chabrier}}$ $\sim 1.5 \times 10^{-8}~yr^{-1}$), the values of $\mathrm{sSFR_{THIMF}}$ are lower (mean value $\sim$ $7.43 \times 10^{-9}~yr^{-1}$). Hence, the trend between the $\mathrm{sM_{Dust}}$ and sSFR shifts while assuming a THIMF with respect to Chabrier IMF owing to these deviations, and we observe lower sSFR values in the top-heavy case.
    \par
\begin{figure*}[ht]
  \centering
    \includegraphics[width=0.9\textwidth]{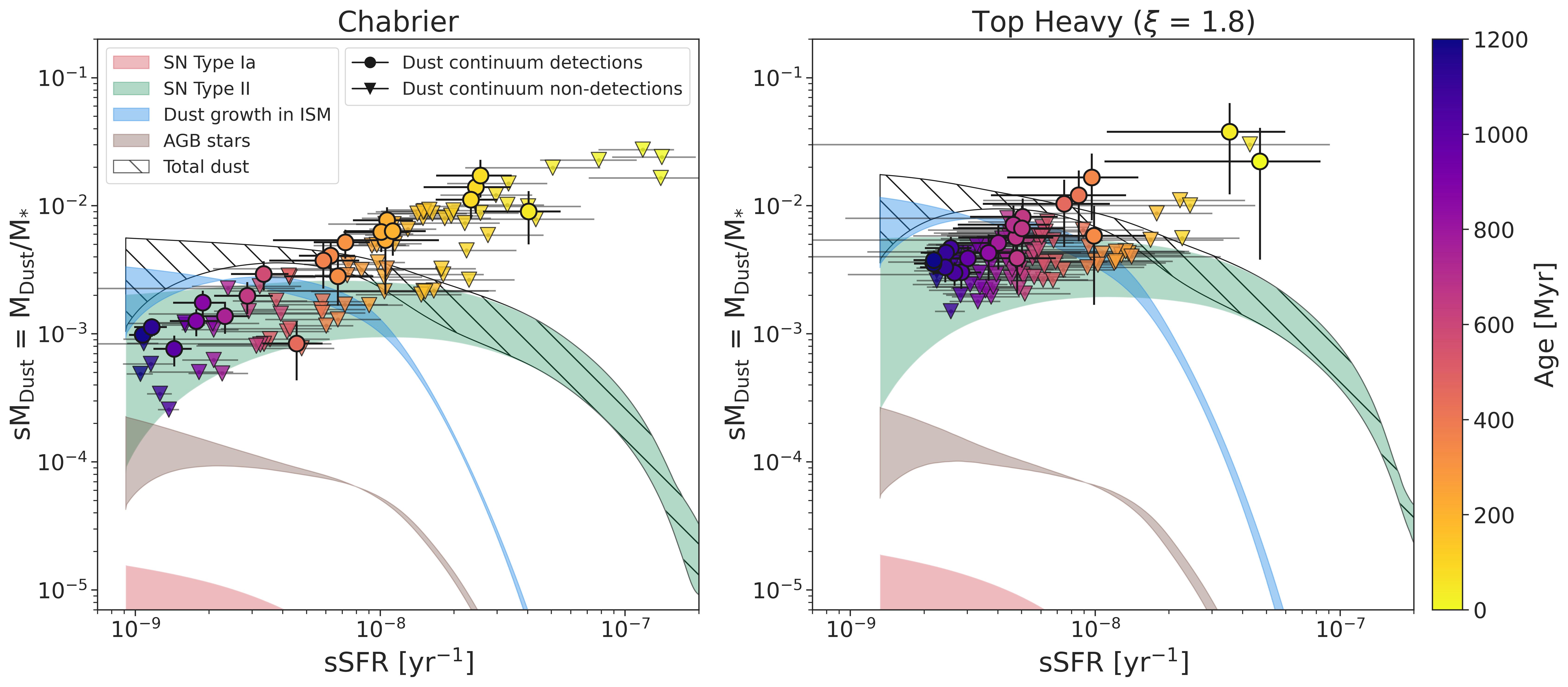}
    \caption{$\mathrm{sM_{Dust}}$ plotted against $\mathrm{sSFR}$ for the ALPINE galaxies assuming a Chabrier IMF (left) and THIMF (right). Circles represent galaxies with detected continuum emission, while inverse triangles mark upper limits on dust mass. Contributions from various dust sources are distinguished by different identifiers. The mass-loading factor ranges from 0 to 2. The overall dust evolution track, which includes all contributions, is shown for the constrained parameters [M$_{\mathrm{Gas, ini}}$, $\eta_{out}$, $\eta_{in}$, $\epsilon_{SN}$] $\sim$ [3.3, 0-2, 0.6, 0.1] and [4.4, 0-2, 0.6, 0.1] for the Chabrier and THIMF cases, respectively, at an age $\sim$ 1200 Myr (hatched region). Green, pink, and brown patches represent contributions from SNII, SNIa, and AGB stars, respectively. For SNII and SNIa, the models range from 25\% dust condensation with $\eta_{out}$ = 0, to 100\% condensation with $\eta_{out}$ = 2. For AGB stars, the models vary with $\eta_{out}$ from 0 to 2. The blue patch shows the contribution of ISM growth to the total dust content, assuming a growth efficiency of 1, varying with $\eta_{out}$ from 0 to 2. Galaxies are color-coded based on the age of their main stellar population.}
    \label{fig:dust_contribution_chabrier}
\end{figure*}

 The extent of the models (i.e., from 25\% of dust condensation with no dust growth in the ISM to 100\% of dust condensation with maximum dust growth efficiency) is represented by the shaded region between the lower and upper bound curves for three predicted ages of the galaxies (i.e., 400, 800, 1200~Myr as representative cases) in Fig.~\ref{fig:DFRD_chabrier}. The dotted curves show the effect of adopting $\eta_{out}=2$, as compared to the case with no outflow (solid patch). For the sake of clarity, we show in Fig.~\ref{fig:DFRD_chabrier} (and in the following, in Fig.~\ref{fig:dust_contribution_chabrier}), only models with condensation fraction larger than 25\%.
    \par In Fig.~\ref{fig:dust_contribution_chabrier}, we show the contribution of each source of dust (i.e., SNIa and SNII, AGB stars, dust growth in ISM) for Chabrier IMF (left) and THIMF (right). Contribution by the dust growth in the ISM is modelled as described by Eq.~\ref{eq:dust_growth}.
    \par For models assuming no outflows, with Chabrier IMF, oldest galaxies with lowest values of sSFR are reproduced with condensation fractions $\sim$ 25\% - 30\% while, assuming a THIMF, higher condensation fractions $\sim$ 50\% are required to reproduce oldest galaxies. For models assuming outflows, with Chabrier IMF, a relatively moderate dust condensation fraction of $\sim 60\%$ is required. The fraction rises to $75-85\%$ in the case of THIMF for galaxies with the same age.
    \par For both Chabrier and THIMF, dust growth plays a significant role for galaxies of intermediate ages (i.e., 300 - 600~Myr), increasing the amount of dust in their ISM of $\sim 60\%$ with respect to the case with no dust growth for condensation fractions $\sim$ 50\%. In the case of a condensation fraction of 5\% for SN (not shown in Fig.~\ref{fig:dust_contribution_chabrier}), dust growth in the ISM becomes the dominant dust production process at sSFR $\approx 2 \times 10^{-8}$~yr$^{-1}$ and sSFR $\approx 3 \times 10^{-8}$~yr$^{-1}$ for Chabrier and THIMF, respectively. For the largest condensation fraction of SN, dust accretion in the ISM starts to dominate at sSFR $\approx 5 \times 10^{-9}$~yr$^{-1}$. The efficiency of dust accretion with the input parameters selected is too low in order to explain the younger galaxies.
    \par We note that dust growth is required to reproduce the specific dust mass in galaxies of intermediate ages (i.e., 300 - 600~Myr), while at older ages models over-predicts the amount of dust formed even for models computed by assuming condensation fraction of 5\%. Since dust in the ISM is more efficient for higher metallicity values (older ages), we may therefore expect another destruction mechanism different from outflow removal and dust destruction from astration, SN shocks and photo-evaporation \citep{Nanni24}, to be at work in destroying the dust grains in the ISM at older ages ($\gtrsim$ 800 Myr). This destruction mechanism can be partially due, for example, to rotational disruption of grains in strong radiation fields \citep{Hoang19}. We also notice that within the uncertainties of the dust mass estimation (a factor of $\sim 3$ lower than \citet{Draine14}), the galaxies of intermediate ages may be compatible with efficient dust production from SNII, without the need of dust growth in the ISM.
 
   \par
\begin{figure*}[ht]
  \centering
    \includegraphics[width=0.9\textwidth]{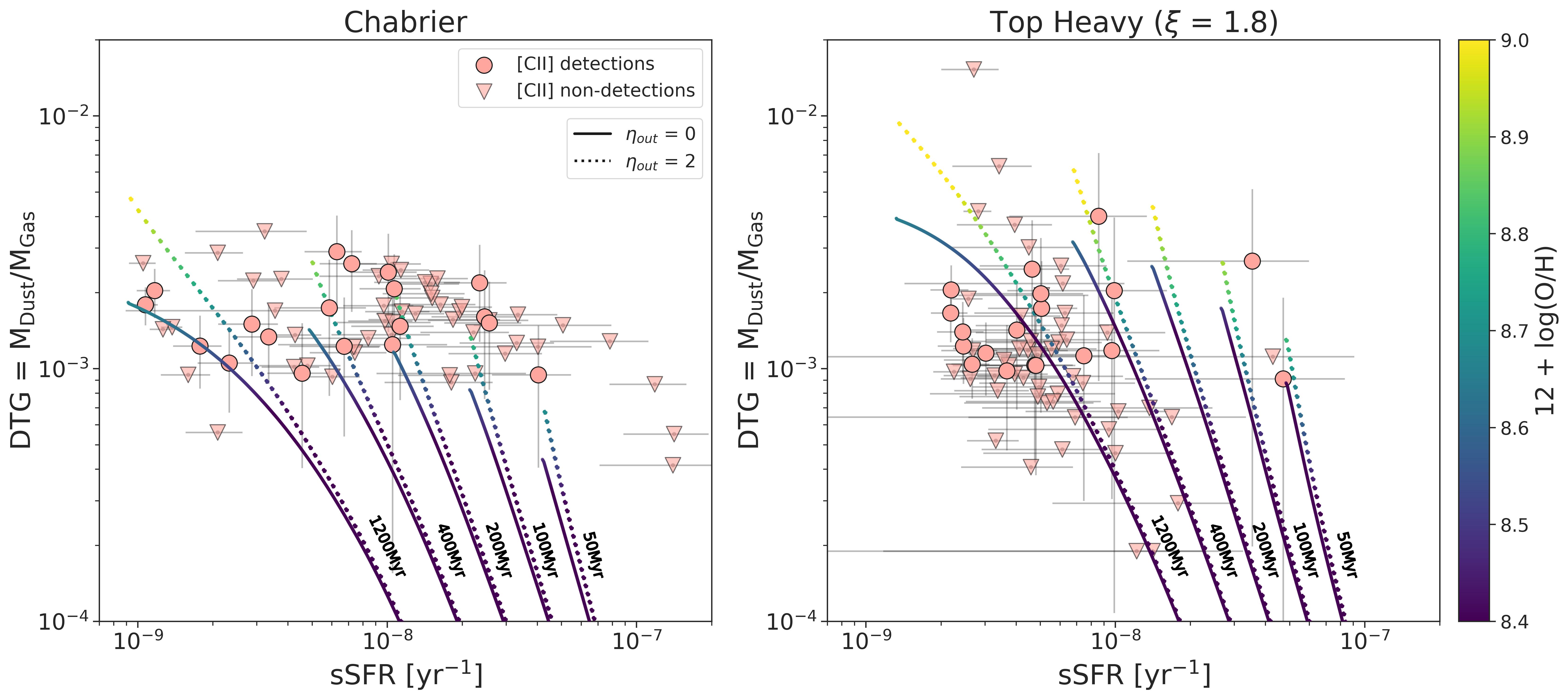}

    \caption{Dust to Gas ratio plotted against sSFR for the ALPINE galaxies assuming a Chabrier (left) and THIMF (right). Only galaxies with [CII] detections are shown. Circles represent galaxies with detected dust continuum, while upper limits on dust mass are indicated by inverse triangles. Evolutionary models are plotted assuming [M$_{\mathrm{Gas, ini}}$, $\eta_{out}$, $\eta_{in}$, $\epsilon_{SN}$] $\sim$ [3.3, 0-2, 0.6, 0.1] and [4.4, 0-2, 0.6, 0.1] for Chabrier and THIMF, respectively. The models are color-coded by predicted metallicity and illustrate the impact of varying the $\eta_{out}$ from 0 (solid curves) to 2 (dotted curves).}\label{fig:dtg_Age}
\end{figure*}

 Furthermore, Fig.~\ref{fig:dtg_Age} shows the observed dust-to-gas (DTG) ratio as a function of the sSFR, along with our predictions from models. We calculate DTG ratio for 68 out of 98 sources which have robust gas measurements as described in Sect.~\ref{subsec:gas_mass}. DTG ratio also provides a solid representation of the baryonic content of the galaxies as in our study, gas and dust measurements are derived independently. As shown in Fig.~\ref{fig:dtg_Age}, the DTG ratio ranges between 5 $\times 10^{-4}$ and 3 $\times 10^{-3}$ in case of a Chabrier IMF, while the points result to be more scattered when assuming a THIMF. Recent observational (\citealt{Donevski20}) and theoretical works employing cosmological simulations (\citealt{Jones24, Palla24}) find around similar DTG ratios.

   \par We are able to reproduce the DTG ratio for the majority of the older galaxies reaching up to 5$\times 10^{-3}$ in case of Chabrier IMF, in agreement with the constraints for $\mathrm{sM_{Gas}}$ and $\mathrm{sM_{Dust}}$ (Fig.~\ref{fig:sMGas_Age} and Fig.~\ref{fig:DFRD_chabrier}, left panels). Assuming a THIMF, we reproduce DTG ratios for younger galaxies and for older galaxies we reach relatively higher DTG ratios, i.e., up to 7 $\times 10^{-3}$.

\section{Discussion}\label{sec:discussion}

   In the following, we discuss the contributions of the diverse mechanisms regulating the gas and dust cycle in the ISM of our galaxies. We also argue on the current limitations of the data and the models in reproducing the observables of youngest and dustier galaxies, as well as on possible solutions. 

   \subsection{Contributions of different processes to the dust cycle}\label{subsec:contributions}
    Numerous studies have investigated the nature and amount of dust in the ISM of galaxies both in the local and high-redshift Universe, suggesting different mechanisms for its production and destruction. AGB stars are considered to be the significant producers of dust in galaxies (\citealt{Clemens10, Valiante11}). However, they are not efficient on short timescales, as those typical of the early Universe (e.g., \citealt{Gall11a,DellAgli19,Tosi23}). Furthermore, dust can survive the SN reverse shock (\citealt{Micelotta16, Marassi19, Slavin20}), or directly grow in the ISM of galaxies by embedding metals from the gas phase (\citealt{Asano13, Hirashita15, Ginolfi18, Aoyama19, Donevski20, Palla24}). Various dust destruction scenarios by SN remnants as a result of uncertainties in the destruction efficiency are suggested where there is inefficient dust destruction in high-temperature gas (\citealt{Vogelsberger19}) and enhanced dust destruction by SN (\citealt{Micelotta18, Hu19, Kirchschlager22}).
    \par In Fig.~\ref{fig:dust_contribution_chabrier}, we observe that, while assuming Chabrier IMF, SNII is the primary source of dust for the older galaxies (i.e., $\gtrsim 600$~Myr). The AGB contribution becomes prominent towards the later stages of their evolution (age $\gtrsim1000$~Myr), and can dominate their total dust production provided that the condensation fraction is less than 25\% for SNII for the case that no dust growth in ISM is present. The contribution from SNIa is always negligible as compared to SNII and AGB together ($\sim$0.001\% of total dust content). Assuming no dust growth, we can reproduce $\sim$ 30\% of galaxies (with age $\gtrsim$ 300~Myr) when $\eta_{out} = 0$ (i.e., assuming no galactic outflow), owing to contribution from SNII and AGB. Considering the case of models with outflows, we can reproduce similar percentage of the sample but assuming higher condensation fractions ($\sim$ 50\%) for SNII. The latter scenario is likely more reliable given the ubiquitous observations of star-formation driven outflows in both local and high-$z$ sources \citep{Romano23,  Mitsuhashi23, Romano24}, and specifically because of the evidence for feedback found in ALPINE (\citealt{Ginolfi20b}). Furthermore, dust growth in the ISM takes over as the major contributor to the dust content where SNII are insufficient with extreme condensation fractions for galaxies with intermediate ages (i.e., 300 - 600~Myr). Similar conclusions have been presented for REBELS survey (z $\sim$ 7, \citealt{Algera24}), where dust growth in ISM is necessary to reproduce dust masses in high-z galaxies.

    \par Fig.~\ref{fig:DFRD_chabrier} shows the gradual trend in the dust cycles as we vary the mass-loading factor. Higher $\eta_{out}$ leads to more outflow of gas which is reflected in the trends for models plotted for sMgas in Fig.~\ref{fig:sMGas_Age} and for $\mathrm{sM_{Dust}}$ in Fig.~\ref{fig:DFRD_chabrier}. With increasing $\eta_{out}$, the contributions from SNII are offset by the dust growth in ISM which takes over to be the primary contributor to the dust content of the older galaxies (age $\gtrsim$ 800~Myr). The dust content in the younger galaxies is denoted by high values of specific dust mass for high values of specific star formation rate. The rapid dust build-up observed in galaxies younger than 500~Myr is not reproducible when considering traditional IMF (i.e., Chabrier IMF). Hence, a different form of IMF is tested and explained in the next section.

    \subsection{Effect of different IMFs}\label{subsec:effect_IMF}       
        \par As discussed in Sect.~\ref{sec:methodology}, we assumed a THIMF introducing a population of short-lived massive stars. The effect of this assumption is reflected in the results represented in Fig.~\ref{fig:dust_contribution_chabrier} (right) where SNII are the primary source of dust for the galaxies with age $\gtrsim$ 300~Myr. Due to the short lives and rapid stellar evolution, the dust build-up observed in the younger galaxies is reproduced with the models reaching larger $\mathrm{sM_{Dust}}$ values in timescales of 400~Myr - 500~Myr. In case of models with no dust growth in ISM (right panel of Fig.~\ref{fig:dust_contribution_chabrier}), galaxies with age $\sim$ 800~Myr are reproduced owing to the contributions from SNII and AGB stars assuming no outflow ($\eta_{out}=0$) and condensation fractions for SNII limited to moderate values ($\sim$25\%). For the galaxies that reach higher values of $\mathrm{sM_{Dust}}$ in shorter timescales ($\sim 300$~Myr), dust growth in ISM is required to reproduce the observations, hence being the primary contributor to the dust content of the galaxies as contributions from SNII and AGB prove to be insufficient. At the same time, if dust growth is considered, we overproduce dust for older galaxies. In case of $\eta_{out} = 2$, dust contribution from SNII reproduced galaxies with higher values of $\mathrm{sM_{Dust}}$ even at older ages only if condensation fraction is $\gtrsim$ 80\% assuming no dust growth.
        \par In Fig.~\ref{fig:dtg_Age} (right), we probe the DTG ratio for the galaxies assuming a THIMF, where the models reproduce the DTG ratio for the younger galaxies which are observed to have higher $\mathrm{sM_{Dust}}$ (Fig.~\ref{fig:dust_contribution_chabrier}, right panel), given that $\mathrm{M_{Gas, ini}}$ is greater than that in case of Chabrier. Assuming a THIMF helps in reproducing higher sM$_\mathrm{Dust}$ values but not higher DTG ratios as THIMF favours rapid dust production from SNII and faster dust growth but at the same time also enhances gas content in the ISM due to the larger mass ejection rates from evolved stellar populations \citep{Palla20}.
        \par Generally, in case of THIMF, it is observed that AGB stars are not contributing significantly in contrast to their contribution when considering the Chabrier IMF. These cumulative trends hint that the nature of dust production and destruction is variational and different in various stages of galactic evolution. In Fig.~\ref{fig:dust_contribution_chabrier} (right), we observe that the THIMF falls short for younger galaxies ($\lesssim$ 100~Myr) with higher values of $\mathrm{sM_{Dust}}$ ($\gtrsim 10^{-2}$). This can be attributed to limited understanding of the mechanisms at play in high-z universe or overestimation of derived parameters from SED fitting of the galaxies, thus hinting at the need of emendation of our models.

    \par For such galaxies, we examined variations in the SED fitting and evolution models in relation to the rest of the sample, with the goal of reproducing their dust content in Fig.~\ref{fig:DFRD_chabrier} and Fig.~\ref{fig:dust_contribution_chabrier}. Our initial tests focused on the possibility that these galaxies possess a larger gas mass, which could explain the values of $\mathrm{sM_{Gas}}$ shown in Fig.~\ref{fig:sMGas_Age}. These values could only be satisfied under the assumption of a THIMF. In contrast, for a Chabrier IMF, increasing the gas mass -- whether by adjusting the initial gas mass ($M_{\rm Gas, ini}$) or increasing the inflow rate via $\eta_{\rm in}$, failed to reproduce the properties of the youngest galaxies ($\mathrm{sM_{Gas}}$) shown in the left panel of Fig.~\ref{fig:sMGas_Age}, as increased gas content in the models would, in theory, reduce the efficiency of dust destruction by SN and dust removal through outflows (see Eq. 6, 8, and 9). However, even under the assumption of maximal condensation fractions of metal into dust grains, the predicted $\mathrm{sM_{Dust}}$ does not reach the observed values.
    \par We further explored the possibility of different star formation history, using a delayed plus burst model (\citealt{Boquien19}). In this scenario, we re-fitted the sources by allowing the fraction of stars to form in the burst, the age of the stellar population, the burst e-folding time, and the burst age to vary as free parameters. Despite these variations, the $\mathrm{sM_{Dust}}$ derived for the youngest galaxies remains too large to be reproduced by the current models.
    \subsection{Comparison with different dust models}
    \par As shown in Figs.~\ref{fig:DFRD_chabrier}, \ref{fig:dust_contribution_chabrier} and discussed in Sect.~\ref{subsec:effect_IMF}, the implementation of a THIMF can alleviate the tension between the observed and predicted amount of dust in the ISM of high-$z$ galaxies. In spite of this, a few sources with high $\mathrm{sM_{Dust}}$ and sSFR still exceed the theoretical expectations, challenging our comprehension of galaxy formation at early times.
    \par Previous attempts of reproducing the observed dust masses in $z>4$ galaxies have faced similar difficulties. \cite{Dayal22} made use of the \textsc{DELPHI} semi-analytical model to explain the dust content in the ISM of 13 galaxies detected in [CII] and dust continuum as part of the REBELS program, whose targets were selected in a comparable way as for ALPINE but at higher redshift ($z>6$) and show similar properties in terms of gas mass, sSFR, and $\mathrm{sM_{Dust}}$ (e.g., \citealt{Dayal22,Palla24}). Their fiducial model (which includes the combined effect of SNII, astration, destruction and removal of dust by shocks and outflows, and a grain growth timescale of $\sim30~$Myr) was able to reproduce the majority of those galaxies, anticipating a typical $\mathrm{sM_{Dust}}\sim10^{-3}$, that is comparable with our predictions (see Figs.~\ref{fig:DFRD_chabrier}, \ref{fig:dust_contribution_chabrier}). On the other hand, the same model struggled to reproduce the youngest, low-mass galaxies in their sample (i.e., sources with the largest sSFR), for which extreme assumptions on dust production (i.e., a rapid grain growth with a timescale below $1~$Myr, no dust destruction by SN shocks or no ejection by outflows) were needed instead, leading to $\sim1~$dex larger $\mathrm{sM_{Dust}}$.
    \par \cite{DiCesare23} investigated the dust build-up process in $z>4$ galaxies by means of cosmological simulations performed with the \textsc{dustyGadget} code \citep{Graziani20}. By taking into account dust production by SN, AGB stars, and grain growth in the ISM, as well as destruction/removal by shocks, astration, and sputtering, their simulations reproduce large amounts of dust ($\mathrm{sM_{Dust}}\sim10^{-2}$) in galaxies with $\mathrm{log_{10}(M_{*}/M_{\odot})}\gtrsim10$. This implies a slight overproduction of dust as compared to our models assuming a Chabrier IMF that can be due, for example, to a shorter timescale of dust accretion in the ISM \citep{Graziani20,DiCesare23}. Still, those models fail in reproducing less-massive galaxies with very large $\mathrm{sM_{Dust}}$ (e.g., \citealt{Witstok22}). Recently, \cite{Palla24} used chemical evolution models to investigate dust evolution in the REBELS survey, exploring the impact of different metallicity enrichment of the ISM on predictions for the dominant dust production and/or destruction mechanisms. They found that both a fast or milder dust build-up evolution is needed to reproduce the dust content or DGR for their sources, with specific dust of mass reaching up to $\mathrm{sM_{Dust}}\sim10^{-2}$. However, they also underpredict the observed amount of dust for younger galaxies with the lowest stellar masses ($\mathrm{log_{10}(M_{*}/M_{\odot})}\sim9$), suggesting that for these objects a THIMF may be adopted. Finally, \cite{Pozzi21} compared the observed dust and gas masses of the continuum-detected ALPINE sources with evolutionary tracks from chemical models by \cite{Calura08,Calura14} for galaxies with different morphological types, namely spiral galaxies and proto-spheroids (the latter being progenitors of local elliptical galaxies). In spite of their large dust masses (up to $\sim1~$dex larger than our estimates given their assumption on low dust temperatures; see also Sect.~\ref{subsec:SED_IMF}), their proto-spheroids models were able to reproduce the majority of the galaxies, even those at the highest $\mathrm{sM_{Dust}}$, reaching values up to $\sim10^{-1.5}$ for tracks with final mass of $\mathrm{log_{10}(M_{*}/M_{\odot})}\sim10$. These results seem to be in contrast with the maximum dust production found in this work, as well as in the above-mentioned models from the literature (e.g., \citealt{Dayal22,DiCesare23,Palla24}). However, it must be noted that the prescriptions about the dust production and/or destruction processes in \cite{Calura08} and in the other models can present substantial differences. For instance, \cite{Calura08} tailored their chemical evolution on galaxies with specific morphological types with varied star formation histories. Furthermore, they do not directly include the effect of outflows in the models, rather they instantaneously eject all the gas out of the galaxy when the energy deposited in the ISM by SN equals the binding energy of the gas \citep{Calura08}. Such a differences could explain the discrepancy with other predictions.
 
\section{Conclusions}\label{sec:conclusions}
\par In this work, we employ chemical evolution models to reproduce the observed gas and dust content in $z\sim5$ main-sequence star-forming galaxies as drawn from the ALMA Large Program ALPINE. We derive physical parameters (SFR, dust mass, stellar mass) and constrain the physical processes influencing gas and dust production and/or consumption in their ISM. Specifically, we investigate the impact of the initial gas mass, outflow mass-loading factor, and inflow parameter on the models, and we test our results on canonical (i.e., Chabrier) or top-heavy IMFs. Our main findings are summarized below:
    \begin{enumerate}
    \item Galactic outflows are essential for reproducing the decrease in the observed gas mass with increasing age while keeping the gas-phase metallicity close to (or below) solar. Outflow efficiencies of the order of $\sim2$ are especially needed for galaxies $\gtrsim300~$Myr old (see Fig. \ref{fig:sMGas_Age}), while lower $\eta_{out}$ are viable for younger sources.
    
    \item Regardless of the adopted IMF, the amount of dust in older galaxies ($\gtrsim600$~Myr) can be reproduced with a major contribution from SNII assuming a condensation fraction of $\gtrsim50\%$, and with no need for dust growth in the ISM. However, in galaxies with intermediate age (i.e., $300-600$~Myr), dust growth starts playing a significant role, allowing for a $\sim60\%$ larger amount of dust. We note that if dust growth in the ISM increases with metallicity as predicted by models \citep[e.g.][]{Asano13}, another destruction mechanism should be at work in order to reproduce the galaxies at older ages. This mechanism may be for example rotational disruption of grains exposed to strong radiation fields. Alternatively, the observed dust masses may be overestimated by a factor of $\approx 50$\% \citep{Burgarella22}.
    
    \item Dust production from SNIa is negligible at all ages for both IMFs. In the case of THIMF, AGB stars also do not play a major role. However, they begin to contribute $\gtrsim 10\%$ for ages $\gtrsim 800$~Myr if assuming a Chabrier IMF (see Fig. \ref{fig:dust_contribution_chabrier}).
    
    \item We find that $65\%$ of our galaxies can be reproduced with a canonical Chabrier IMF. The fraction increases to $93\%$ when adopting a THIMF. A flatter slope of the IMF could thus help in alleviating the tension between observations and models for galaxies at the first stages of their dust cycle, although it is still not enough to reproduce the youngest and dustier sources of our sample (sSFR $\gtrsim 10^{-8}~yr^{-1}$, $\mathrm{sM_{Dust}}\gtrsim 10^{-2}$).
   \end{enumerate}
   
   \par Our work highlights the importance of chemical evolutionary models while probing the baryon cycle of primordial galaxies. Although our results seem to suggest the need for a THIMF to allow for a fast production of dust in the youngest sources, further investigation is essential. In this regards, follow-up observations with ALMA are required to sample the peak of the FIR SED of our galaxies, allowing for a better constrain of the dust mass estimates. Furthermore, we are currently taking advantage of a completed JWST/NIRSpec IFU program (ID: 3045, PI: A. Faisst) which collected observations of 18 ALPINE galaxies at kpc scales in the rest-frame optical regime, covering several emission lines including [OIII] or H$\alpha$. This program will allow us to put robust constraints on stellar masses, SFHs, and metallicities of our galaxies, and to calibrate our models for a more in-depth and thorough description of galaxy formation at early times.

\begin{acknowledgements}
We warmly thank the referee for her/his useful comments and suggestions that greatly improved the quality of our paper. P.S., A.N., and M.R. acknowledge support from the Narodowe Centrum Nauki (UMO-2020/38/E/ST9/00077). M.R. acknowledges support from the Foundation for Polish Science (FNP) under the program START 063.2023. D.D. acknowledges support from the National Science Center (NCN) grant SONATA (UMO-2020/39/D/ST9/00720).  J. and K.M. are grateful for the support from the Polish National Science Centre via grant UMO-2018/30/E/ST9/00082. J. acknowledges support from the European Union (MSCA EDUCADO, GA 101119830 and WIDERA ExGal-Twin, GA 101158446). M.B. gratefully acknowledges support from the ANID BASAL project FB210003 and from the FONDECYT regular grant 1211000. This work was supported by the French government through the France 2030 investment plan managed by the National Research Agency (ANR), as part of the Initiative of Excellence of Université Côte d’Azur under reference number ANR-15-IDEX-01. M.H. acknowledges support from the Polish National Science Center (UMO-2022/45/N/ST9/01336). E.I. acknowledges funding by ANID FONDECYT Regular 1221846. G.E.M. acknowledges the Villum Fonden research grant 13160 “Gas to stars, stars to dust: tracing star formation across cosmic time,” grant 37440, “The Hidden Cosmos,” and the Cosmic Dawn Center of Excellence funded by the Danish National Research Foundation under the grant No. 140.
\end{acknowledgements}

%
%
\bibliographystyle{aa} 
\bibliography{aanda.bib} 




\begin{appendix} 

\section{Treatment of upper limits and homogeneity of the sample}\label{app:upper_limits}
In our study, we investigate both [CII] and dust continuum detected and non-detected galaxies from the ALPINE sample and compare them with our models. Indeed, although detected sources offer more robust constraints on the gas and dust evolution, non-detections typically extend to lower stellar masses and SFRs, providing fundamental information on the baryon cycle in galaxies at the low-mass end of the main-sequence. We stress here that the inclusion of non-detections in this work does not affect the homogeneity of our sample (e.g., \citealt{Schaerer20,Romano22}).

\cite{Schaerer20} made a distinction between galaxies in ALPINE detected both in [CII] and dust-continuum, and those detected only in [CII]. In their work (refer to their Fig. 4), they examined the relationship between [CII] luminosity and total star formation rate ($\mathrm{SFR_{total}}$, UV + IR). For the dust continuum-detected sources, the obscured SFR was derived using a relation between monochromatic luminosity at 158~$\mu m$ and the IR luminosity (\citealt{Bethermin20}). For the non-detected sources, the IRX-$\beta_{FUV}$ relation obtained from stacking of ALPINE sources (\citealt{Fudamoto20}) was employed. Notably, this analysis revealed that after correcting for dust-obscured star formation, both the continuum-detected and non-detected galaxies agree with local [CII]-$\mathrm{SFR_{total}}$ relations (\citealt{DeLooze14, Lagache18}), showing no intrinsic differences between these subsets. Importantly, dust continuum non-detections tend to have lower SFRs, subject to the ALMA sensitivity threshold rather than indicating they are outliers from the dust continuum-detected population. \cite{Bethermin20} highlighted this point by performing a stacking analysis across continuum detected and non-detected sources in bins of [CII] luminosity, finding that all galaxies were within the scatter of the local [CII]-SFR relations. Similarly, \cite{Romano22} analyzed [CII]-undetected galaxies by stacking their ALMA spectra at the optical position of the sources. They found that the corresponding stacked [CII] emission follows the above-mentioned [CII]-SFR relations, suggesting that [CII] non-detections are drawn from the same population of [CII]-detected galaxies, just being fainter in [CII] due to the lower SFRs.

\par Further support for this result comes from \cite{Burgarella22}, who re-examined the IRX-$\beta_{FUV}$ relation using a stacked template for the same galaxies. They found that this approach reduced the scatter in $\beta_{FUV}$ values reported by \cite{Fudamoto20}, attributing this reduction to more accurate IR luminosity estimates derived from the stacked template. This supports the reliability of using stacked templates to estimate dust content in continuum non-detected galaxies, as well (see also Appendix~\ref{app:stacked_template}).

\par In our analysis, we observe a shift in the median stellar mass and SFR between dust-continuum detected and non-detected galaxies. The Kolmogorov-Smirnov (KS) test yields a low p-value ($\sim$$10^{-5}$), indicating a significant difference between these two populations. However, when comparing normalized quantities, such as specific star formation rate (sSFR) and specific dust mass ($\mathrm{sM_{Dust}}$), the KS test returns higher p-values (0.2 and 0.8, respectively), suggesting that both populations are consistent with being part of a homogeneous sample when examined through these normalized physical parameters. This homogeneity is reflected in the DFRD presented in Fig.~\ref{fig:DFRD_chabrier}.

\section{Reliability of stacked IR template}\label{app:stacked_template}
To compensate the shortage or complete absence of data coverage in the IR regime of our galaxies, \cite{Burgarella22} built an IR composite template from a sub-sample of ALMA-detected objects (including 20 dust continuum-detected galaxies from our sample), widely covering the peak of the FIR SED. They showed that the template is not biased against peculiar sources or outliers, but rather is valid for samples of galaxies with similar selection criteria (see \citealt{Burgarella22} for more details).

\begin{figure*}[h]
  \centering
    \includegraphics[width=0.9\textwidth]{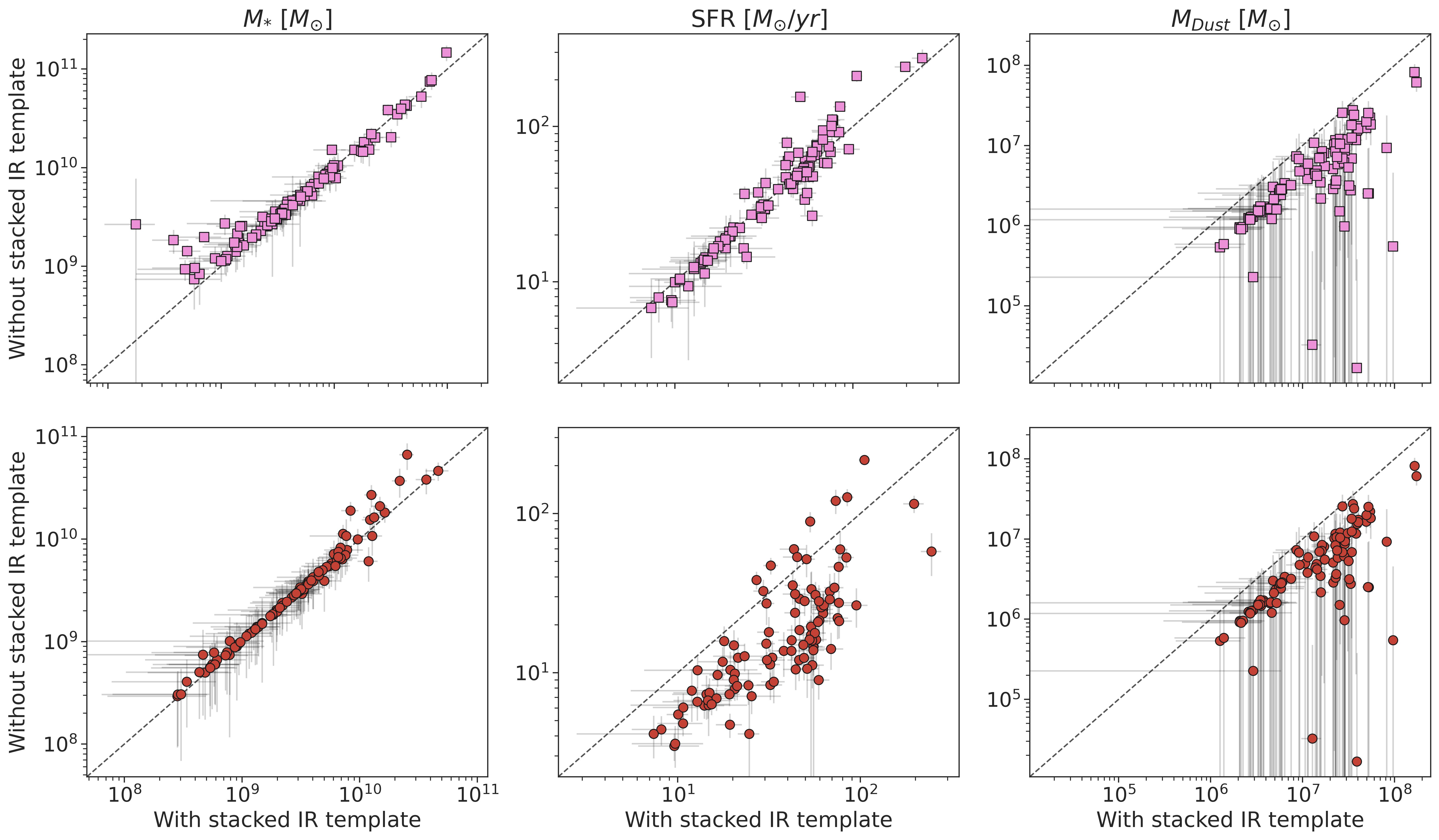}
    \caption{Comparison of physical parameters ($\mathrm{M_{*}}$, SFR, $\mathrm{M_{Dust}}$) of ALPINE galaxies derived with and without the stacked IR template in \textsc{cigale}. Top and bottom panels denote parameters derived from SED fitting assuming a Chabrier and a THIMF, respectively.}
    \label{fig:stack_vs_nostack}
\end{figure*}

To further ensure that the inclusion of the stacked template does not significantly affect our conclusions, we performed SED fitting of our galaxies without including the IR template, and compared the corresponding physical properties with those obtained in our work (i.e., including the template). As shown in Fig.~\ref{fig:stack_vs_nostack}, the parameters obtained by using the stacked template (both in the case of Chabrier and THIMF) are better constrained, especially in the derivation of dust masses (more affected by a poor IR coverage). This is also confirmed by a mock analysis with \textsc{cigale}, made to test the reliability of our SED fitting process. In brief, for each galaxy, an artificial (mock) photometric catalog is created by using its best-fit model from \textsc{cigale}, and adding as error a randomly-derived value from a Gaussian distribution with the same standard deviation as the observed uncertainty (e.g., \citealt{Boquien19}). These simulated data are then fitted with the same input quantities we adopted for SED fitting (see Sect.~\ref{sec:methodology}), to derive physical parameters of the artificial galaxies that we compared with the input (unperturbed) ones. We made this for both the cases with and without the stacked IR template. As a result, we found that simulated and input parameters lie mostly along the $\mathrm{1:1}$ relation, with a strong positive correlation (i.e., $\mathrm{r^2>0.85}$, with $\mathrm{"r"}$ being the Pearson's correlation coefficient) in all cases. However, when no composite template is used, the scatter across the relation increases, especially in the case of dust masses. This ensures us that the inclusion of the stacked IR template from \cite{Burgarella22} in our SED-fitting analysis provides the best results to our sample. 

\section{Schematic for estimation of optimal model parameters}\label{app: fitting_scheme}

\begin{figure*}[h]
  \centering
    \includegraphics[width=0.9\textwidth]{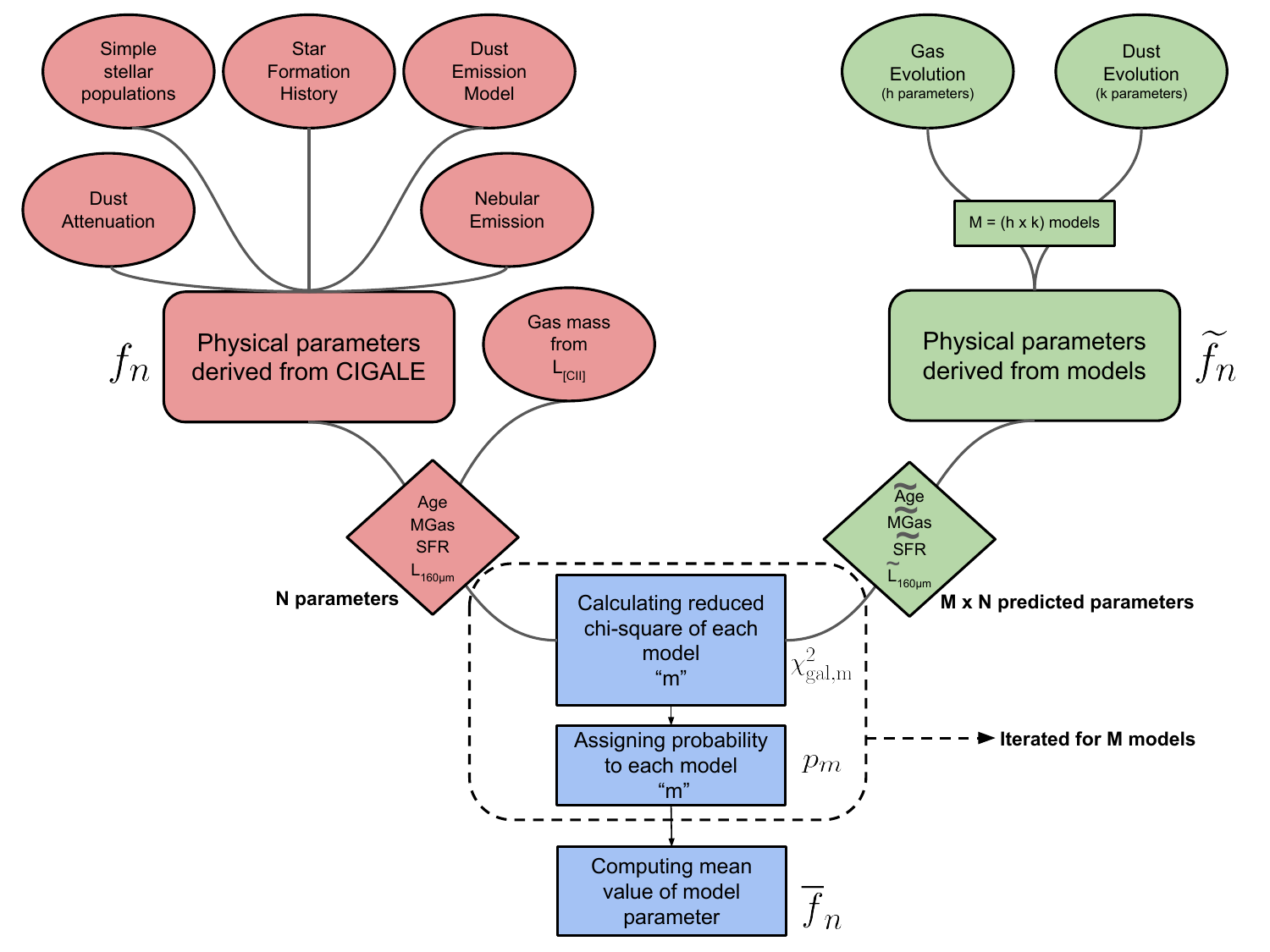}
    \caption{Visual representation of the fitting procedure presented in Sect.~\ref{subsec:best_fit}. Steps shaded in red denote derivation of $N$ physical parameters from SED fitting with \textsc{cigale}. Green steps denote derivation of $N$ predicted physical parameters from $M$ chemical evolutionary models. Blue steps denote statistical analysis to derive the optimal value of our parameters.}
    \label{fig:fitting_scheme}
\end{figure*}

In Fig.~\ref{fig:fitting_scheme}, we schematize the procedure employed to estimate the optimal values of the model parameters (${M_\mathrm{Gas, ini}}$, $\eta_{in}$, and $\epsilon_{SN}$). We first derive $N$ physical parameters such as age, gas mass, SFR, and dust luminosity at 160~$\mu$m from SED fitting using \textsc{cigale}. Simultaneously, for $M$ models (generated from a grid of $h\times k$ parameters, as taken from the gas and dust evolution, respectively), we derive $M\times N$ predicted physical parameters. We perform a reduced chi-square minimization, from which we calculate the probability ($p_{m}$) for each model, where $m$ represents an individual model, ranging from 1 to $M$. The mean value of each model parameter ($\overline{f}_{n}$) is then determined by weighting each model according to its respective probability. This procedure is discussed in detail in Sect.~\ref{subsec:best_fit}.

\end{appendix}

\end{document}